\documentclass[acmtog,screen,review=False]{acmart}

\setcopyright{none}
\renewcommand\footnotetextcopyrightpermission[1]{}

\AtBeginDocument{\providecommand
\BibTeX{{\textrm B\kern-.05em{\textsc i\kern-.025em b}\kern-.08em\TeX}}
}

\setcopyright{acmcopyright}
\copyrightyear{2022}
\acmYear{2022}
\acmDOI{TBD}

\acmJournal{TIST}
\acmVolume{0}
\acmNumber{0}
\acmArticle{0}
\acmMonth{0}

\usepackage{tabularx}
\usepackage{float}

\graphicspath{{figures/}}

\begin{document}

\title{Adaptive Agents and Data Quality in Agent-Based Financial Markets}

%% The "author" command and its associated commands are used to define
%% the authors and their affiliations.
%% Of note is the shared affiliation of the first two authors, and the
%% "authornote" and "authornotemark" commands
%% used to denote shared contribution to the research.
\author{Colin M. Van Oort}
\email{cvanoort@mitre.org}
\orcid{0000-0002-2804-9696}
\affiliation{
    \institution{The MITRE Corporation}
    \streetaddress{7515 Colshire Drive}
    \city{McLean}
    \state{Virginia}
    \country{USA}
    \postcode{22102-7539}
}

\author{Ethan Ratliff-Crain}
\email{ethan.ratliff-crain@uvm.edu}
\orcid{}
\affiliation{
    \institution{The University of Vermont}
    \streetaddress{85 South Prospect Street}
    \city{Burlington}
    \state{Vermont}
    \country{USA}
    \postcode{05405}
}

\author{Brian F. Tivnan}
\email{btivnan@mitre.org}
\orcid{0000-0001-7079-8183}
\affiliation{
    \institution{The MITRE Corporation}
    \streetaddress{7515 Colshire Drive}
    \city{McLean}
    \state{Virginia}
    \country{USA}
    \postcode{22102-7539}
}

\author{Safwan Wshah}
\email{Safwan.Wshah@uvm.edu}
\orcid{0000-0001-5051-7719}
\affiliation{
    \institution{The University of Vermont}
    \streetaddress{85 South Prospect Street}
    \city{Burlington}
    \state{Vermont}
    \country{USA}
    \postcode{05405}
}
%%
%% By default, the full list of authors will be used in the page
%% headers. Often, this list is too long, and will overlap
%% other information printed in the page headers. This command allows
%% the author to define a more concise list
%% of authors' names for this purpose.
\renewcommand{\shortauthors}{Van Oort, et al.}

\begin{abstract}
    We present our Agent-Based Market Microstructure Simulation (ABMMS), an Agent-Based Financial Market (ABFM) that captures much of the complexity present in the US National Market System for equities (NMS).
    Agent-Based models are a natural choice for understanding financial markets.
    Financial markets feature a constrained action space that should simplify model creation, produce a wealth of data that should aid model validation, and a successful ABFM could strongly impact system design and policy development processes.
    Despite these advantages, ABFMs have largely remained an academic novelty.
    We hypothesize that two factors limit the usefulness of ABFMs.
    First, many ABFMs fail to capture relevant microstructure mechanisms, leading to differences in the mechanics of trading.
    Second, the simple agents that commonly populate ABFMs do not display the breadth of behaviors observed in human traders or the trading systems that they create.
    We investigate these issues through the development of ABMMS, which features a fragmented market structure, communication infrastructure with propagation delays, realistic auction mechanisms, and more.
    As a baseline, we populate ABMMS with simple trading agents and investigate properties of the generated data.
    We then compare the baseline with experimental conditions that explore the impacts of market topology or meta-reinforcement learning agents.
    The combination of detailed market mechanisms and adaptive agents leads to models whose generated data more accurately reproduce stylized facts observed in actual markets.
    These improvements increase the utility of ABFMs as tools to inform design and policy decisions.
\end{abstract}

\maketitle

\section{Introduction}\label{sec:intro}
Decades of market microstructure research have shown that the mechanics of trading meaningfully impact price formation~\cite{ohara1997market, madhavan2000market, hasbrouck2007empirical}.
Price formation in agent-based financial markets (ABFMs) is also influenced by market microstructure, thus, ABFMs that fail to adequately capture market microstructure mechanisms present in their target systems may observe divergences in behaviors and outcomes.

The ecology of agents that populate an ABFM is just as important as the market infrastructure that mediates their interactions~\cite{lebaron2001builder}.
Simple agents, such as Zero intelligence (ZI) agents~\cite{gode1993allocative}, are commonly used to understand baseline characteristics of markets~\cite{smith2003statistical, farmer2005predictive, lebaron2006agent, cont2010stochastic, gould2013limit}.
However, simple agents do not exhibit the level of heterogeneity or adaptability seen in authentic market participants.

In real markets, it is common for short term trading strategies to lose effectiveness over time, a phenomenon referred to as alpha decay~\cite{di2016alpha}.
By some estimates, short term strategies take 3 to 7 months to develop and remain effective for 3 to 4 months~\cite{schmerken2008alpha}.
Since the average development duration is longer than the average strategy lifetime, we might expect the population of short term strategies to have a high turnover rate.
This high turnover rate may be a mechanism driving non-stationary trading dynamics.
It also indicates that strategy adaptation is a critical attribute of successful market participants, and that static strategies may be poorly suited for realistic ABFMs.

The use of adaptive strategies in ABFMs can promote agent specialization, leading to emergent heterogeneity.
Agent heterogeneity can contribute to financial market resilience~\cite{bookstaber2016toward}, thus emergent heterogeneity driven by adaptive agents could improve the stability of ABFMs.
Additionally, agent adaptability is critical to realize economic rationality in non-trivial ABFMs~\cite{vives1993fast, hammond1997rationality, marwala2017rational}.
Economically rational agents avoid using trading strategies that lead to financial ruin.
Thus, when agents with inflexible strategies encounter unfavorable market conditions they may be forced to exit the market.
Excessive attrition can lead to complete failure of a simulated market.
Observed deviations from the Efficient Markets Hypothesis~\cite{malkiel1970efficient, brown2011efficient} and the rise of the Adaptive Markets Hypothesis~\cite{lo2004adaptive} indicate a growing realization of the importance of agent adaptability in financial markets.

In this paper we present our Agent-Based Market Microstructure Simulation (ABMMS), an ABFM with realistic market mechanisms and agent adaptability as core design principles.
We evaluate ABMMS under different configurations to determine the impacts of market fragmentation and adaptive agents on the quality of generated data.
Our evaluation procedure is built using stylized facts and analytical methods developed by the econometrics, market microstructure, and ABFM communities.
ABMMS is able to produce market data that mimics the form and features seen in genuine market data products.
Data generated by ABMMS conforms to several stylized facts of asset prices, along with other observed properties of authentic market data, and thus may be more suitable to inform system design or policy than simpler ABFMs.

\section{Related Work}\label{sec:related-work}
\subsection{Market Infrastructure in the National Market System}\label{subsec:nms-infra}
ABMMS aims to reproduce aspects of the US National Market System for equities (NMS), and many design decisions were based on this choice of target system.
To provide the appropriate context for understanding our model, we summarize the market infrastructure present in the NMS and indicate references with additional details.

Trading in the NMS occurs in a fragmented market that consists of 16 securities exchanges, which manage several continuous double auctions (CDAs) to support trading in a set of assets (equity securities, exchange traded funds, etc).
A CDA allows traders to submit orders to buy (bid) or sell (offer) at any time, and processes those orders immediately upon receipt.
Orders that cannot be fulfilled instantly are collected in a Limit Order Book (LOB).
Almost all CDAs prioritize order execution based on price and time, i.e. price-time priority, though some may use additional attributes.
For additional details regarding CDAs or LOBs, see one or more of ~\citet{smith2003statistical, gould2013limit, abergel2016limit}, and ~\citet{friedman2018double}.

The 16 exchanges that form the NMS are housed within at least four data centers in northern New Jersey~\cite{tivnan2020fragmentation}.
These data centers are connected by an Communication Network (CN) that is implemented with a combination of fiber optic technology and wireless alternatives~\cite{anova2021llfc}.

The use of continuous trading mechanisms causes a race to react any time new public information is released.
All known communication technologies are limited by the speed of light, which guarantees the existence of propagation delays on each leg of an CN, an average of 100 $\mu$s based on the current configuration of the NMS~\cite{tivnan2020fragmentation}.
Optimized trading algorithms on specialized hardware may only take between tens of ns to a few $\mu$s to react to incoming messages.  % TODO: Maybe add argument based on IPS?
The combination of these three properties means that the propagation delays imposed by the topology and geometry of the CN can have an immense impact on trading outcomes.
See Section 3 of ~\citet{tivnan2020fragmentation} for additional details regarding the organization of the NMS\@.

Beyond the mechanical details mentioned above, the regulatory environment that surrounds the NMS plays an important role in shaping the market infrastructure and agent behaviors.
% TODO: Consider adding a sentence about SIPs and a sentence about trade-through protection
Readers interested in understanding key NMS regulations should refer to Appendix 3 from \citet{tivnan2020fragmentation} for an overview, or the regulation itself for details~\cite{sec2005regnms}.

\subsection{Market Infrastructure in Prior ABFMs}\label{subsec:model-infra}
Many ABFMs implement a simple infrastructure that allows clear emphasis to be placed on specific elements, while also reducing computational costs and allowing for rapid experimentation.
For example, ~\citet{wah2017welfare} study a population of heterogeneous agents trading a single asset in a single continuous double auction (CDA).
The simplicity of the infrastructure places the emphasis on the heterogeneous agents, the quality of their interactions, and differences in their outcomes.

It is possible to model financial markets without agents or an explicit market microstructure.
Equation Based Models (EBMs) boil down all activity to a set of mathematical equations, commonly differential equations, that describe macro-level quantities, such as asset prices~\cite{sabzian2019theories}.
However, abstracting away these details can restrict or eliminate the possibility of emergent phenomena, greatly reducing the expressiveness of a model.

Early modeling efforts focused on simpler market architectures, such as Walrasian auctions and dealer markets~\cite{ohara1997market, hasbrouck2007empirical}.
But most modern equity markets feature a fragmented CDA with trading activity distributed across multiple locations.

Some have used ABFMs to investigate the impacts of market fragmentation, usually focusing on the simplest case involving two auctions~\cite{wah2016latency, duffin2018agent, baldauf2021fragmented}.
To simulate fragmented markets these ABFMs must account for communication latency, otherwise the fragmentation would not have a material impact on trading activity.
When modeling fragmented markets, it is common to implement one or more securities information processors (SIPs)~\cite{tivnan2017sip}.
SIPs serve as data aggregators that disseminate important signals to keep prices synchronized in a fragmented market, such as the National Best Bid and Offer (NBBO), an indicator of market wide best price.

In addition to market fragmentation, which occurs at the level of financial exchanges, some have explored ABFMs that capture the interactions that occur in multi-level markets containing many interconnected financial systems, such as equity markets, options markets~\cite{ecca2008modeling}, brokerages, and banks~\cite{bookstaber2018agent}.

Speed can be a deciding factor in the competition of trading strategies, especially in market systems with continuous auction mechanisms.
An often underappreciated element of this competition are response delays, the time it takes a trading strategy to ingest an incoming market message and issue an appropriate response.
These response delays are often so small that they are assumed to have minimal impact on trading outcomes, and thus are not implemented in many ABFMs.
However, since many aspects of the race for speed have been commoditized (e.g., colocation, wireless communication channels, specialized computing hardware, etc.) the microseconds that can be shaved via software optimization can have serious impacts~\cite{rollins2020trading}.

There are an endless number of market infrastructure details that can impact trading processes, and should be captured in detailed models.
However, in this work we focus exclusively on a stock market with detailed implementations of market fragmentation, communication infrastructure, auction mechanisms, and adaptive agents.

We applaud recent work in this area, which has attempted to address aspects of market infrastructure realism or trading agent adaptability~\cite{jacob2016rock, wang2017spoofing,byrd2019abides,dicks2022simple,gao2022high,gao2022understanding}, though we believe that ABMMS implements a unique set of features that is not entirely present in any existing model.

\subsection{Adaptive Agents}
Mechanisms for adaptive agents can be classified as active or passive~\cite{lebaron2011active}.
Active learning involves intentional change of an agent's strategy, while passive learning occurs via the accumulation of wealth by more effective strategies over time.
We direct our interest towards active learning mechanisms, due to recent advances in the field of machine learning as well as the potential relationship between active learning and economic rationality~\cite{vives1993fast}.

Active adaptive agents can feature two types of strategies, fixed or free form.
Fixed strategies cover a single qualitative class of behaviors and tune a set of parameters in response to changing market conditions.
Despite their ability to modify certain aspects of their behavior, such as interaction frequency or pricing beliefs, fixed strategies cannot spontaneously adopt qualitatively distinct strategies.
On the other hand, free form strategies are able to implement two or more classes of behavior, and perhaps even develop new strategies on the fly.
Since the behavior of fixed strategies is more constrained than their free form counterparts, they tend to be simpler to develop and understand.

\subsubsection{Fixed Strategy Agents}
The ZI agents introduced by ~\citet{gode1993allocative} were simple and broadly applicable, which lead to a proliferation of applications and sparked a vein of research that has been developed for decades.
ZI Plus (ZIP) agents extend the ZI recipe by developing pricing beliefs based on the bid and offer prices that lead to trades~\cite{cliff1997zero, cliff1997minimal, preist1998adaptive, cliff2018bse}.
The agents created by ~\citet{gjerstad1998price} (GD) have a similar structure to ZIP agents, they develop price belief functions based on quotes and trades.
However, GD agents take actions that greedily maximise surplus, where as ZIP agents do not directly optimize profits.
By covering some pathological edge cases in the GD algorithm, Modified GD (MGD) agents~\cite{tesauro2001high} avoid excessive volatility and outperform their predecessor.
The GDX strategy~\cite{tesauro2002strategic} also builds on the pricing belief functions seen in GD agents, but accounts for future rewards via dynamic programming.
This forward-looking optimization promotes longer-term strategies with more interesting behavior.
Adaptive Aggressiveness (AA) agents~\cite{vytelingum2008strategic} combine price belief functions with an aggression function that allows them to strategically account for their ``desire to trade''.
Taking a slightly different approach, Assignment Adaptive (ASAD) agents~\cite{stotter2014behavioural} use a relatively simple strategy that is less adaptive in some ways than ZIP agents, but explicitly accounts for the information contained in an agent's submitted orders.
ASAD agents can generate interesting dynamics, especially when reacting to exogenous price shocks in a homogeneous strategy space, but are generally outclassed by ZIP agents when in direct competition.

This line of research has created several relatively simple agents that combine domain knowledge with basic machine learning and optimization techniques, resulting in adaptive, but fairly restricted strategies.
Through the use of more advanced machine learning techniques, removing imposed strategy structure, and allowing for greater strategy complexity, we can create agents that develop qualitatively distinct strategies.

\subsubsection{Free Form Strategy Agents}
Free form strategies are constructed around a behavior adaptation mechanism, commonly implemented using machine learning, that allows the agent to respond appropriately to changing market conditions.

Supervised learning techniques, in the form of imitation learning, can replicate observed patterns in order flow data~\cite{le2018deep, sirignano2019universal, wray2020automated}.
However, agents built with imitation learning tend to regurgitate observed behaviors, and thus have little ability to respond to market conditions that were not observed during training or to generate new strategies.
Generative Adversarial Networks (GANs) are able to create realistic looking streams of order flow in a similar manner to imitation learning based methods.
GANs may be better than imitation learning at generating novel content, due to the adversarial learning mechanism, but still lack a mechanism necessary for effective generalization~\cite{li2020generating}.

One of the most obvious learning signals present in financial markets is profit.
Profit motive is relied on as one of the fundamental forces in financial markets, and it makes intuitive sense to train trading strategies with it.
Two classes of algorithms are particularly effective at deriving appropriate behavior from arbitrary reward signals: meta-heuristic search and reinforcement learning.
Both meta-heuristic search~\cite{subramanian2006designing, hu2015application} and reinforcement learning~\cite{schvartzman2009stronger, deng2016deep} have been applied repeatedly, and with varying degrees of success, to the learning of trading strategies.

Many traditional applications of meta-heuristic search and reinforcement learning focused on narrowly defined problems, and did not emphasize the ability to adapt to dynamic environments.
The rise of meta-learning, commonly described as learning to learn, has greatly improved the ability of machine learning models to learn from, and adapt to, more broadly defined problems~\cite{hospedales2020meta}.
Trading agents developed using meta-learning techniques, such a hierarchical reinforcement learning~\cite{talla2016hierarchical} or meta-learned evolutionary strategies~\cite{sorensen2020meta}, often learn more quickly, display higher peak performance, and handle new market conditions more gracefully than agents developed with traditional techniques.

\subsection{Model Examination}\label{subsec:model-exam}
For an ABFM to be useful to policy makers and system designers, it must satisfy three properties.
First, the model must align with the system that it is intended to influence.
Second, the model must provide useful insights into that target system.
Third, the model must garner a certain amount of trust from policy makers and designers.

\begin{figure*}
    \centering
    \includegraphics[width=\textwidth]{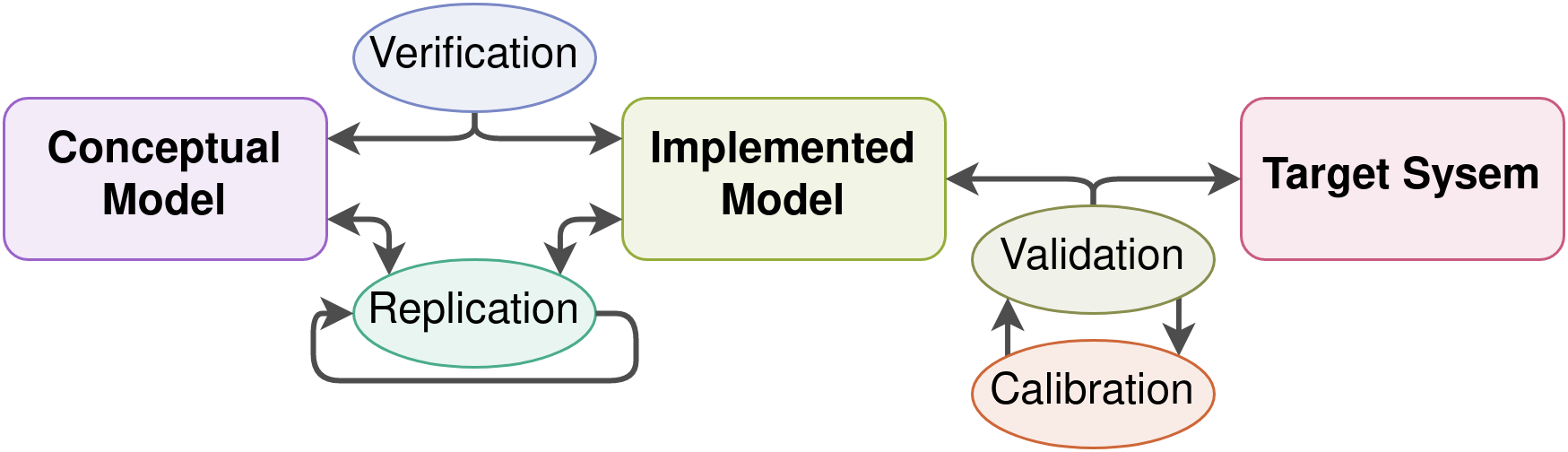}
    \caption{
        The model development process, which includes ABFM development, involves three entities (the conceptual model, implemented model, and target system) connected by four processes (verification, validation, calibration, and replication).
    }
    \label{fig:abm_pipeline}
\end{figure*}

Figure~\ref{fig:abm_pipeline} summarizes the model development pipeline, including ABFM development, which is driven by four processes that ensure quality and consistency: verification, validation, calibration, and replication~\cite{xiang2005verification, rand2006verification, arifin2015verification}.
During verification, an implemented model is compared and contrasted with a conceptual model.
Good software testing and debugging are core verification tasks, though visual inspection of model outputs and other similar actions also play a role.
Validation compares an implemented model with the target system via iterative calibration, which involves tuning free model parameters so that data generated by the model resembles data from the target system.
Comprehensive validation and verification, along with clear communication, establish a baseline level of trust in a model.
Replication, which is a collection of tasks ranging from running code provided by the creators of a model to complete re-implementation, can further bolster the reputation of a model.
The primary goal of replication is to ensure that the outputs of the model display the advertised properties, and are not the result of spurious factors.

Model validation can be driven by data collected from the target system, stylized facts that have been developed based on quantitative observation of the target system, or other forms of distilled knowledge.
In most cases, this decision is based on data availability.
For example, it is prohibitively costly to obtain high frequency data from all of the exchanges in the NMS\@.
Like many who have come before us~\cite{ghoulmie2005heterogeneity, bookstaber2016toward, mcgroarty2019high}, we validate our model using stylized facts of asset price time series~\cite{cont2001empirical}, order book metrics~\cite{paddrik2017effects}, and dislocations~\cite{tivnan2020fragmentation}, instead of depth-of-book data.

\section{Methods}\label{sec:abmms-methods}
\subsection{Market Infrastructure in ABMMS}\label{subsec:abmms-infra}
We developed ABMMS, a highly configurable ABFM that targets the US National Market System for equities (NMS), to investigate the impacts of market microstructure and adaptive agents.
ABMMS emphasizes the explicit representation of many market microstructure elements, starting with a realistic communication network (CN).
The CN consists of a queue of in-flight messages and a topology that those messages travel over.
The topology is an undirected graph with weighted edges, where nodes are data centers, edges are communication channels, and edge weights are deterministic propagation delays.
Messages are routed based on the shortest weighted path, identified via Dijkstra's algorithm.
Exponential noise with a mean of 5 $\mu$s is added to the propagation delay to simulate latency jitter and other stochastic delays.

\begin{figure*}
    \centering
    \includegraphics[width=\textwidth]{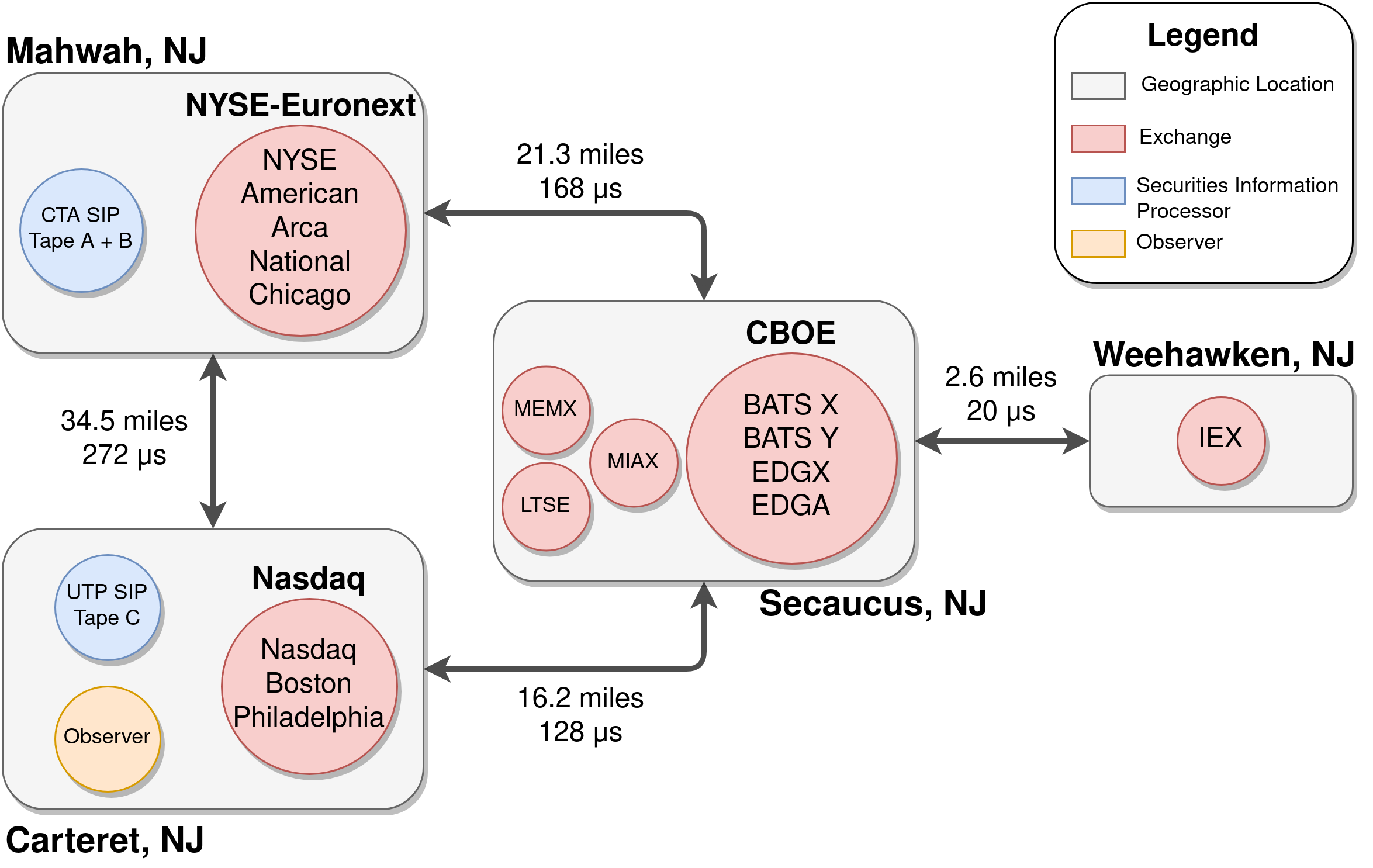}
    \caption{
        A visual summary of the default configuration of ABMMS\@.
        The topology and propagation delays are adopted from ~\citet{tivnan2020fragmentation}, with four data centers distributed across northern New Jersey.
        The choice of 16 exchanges and 2 SIPS is based on our understanding of the NMS in early 2021.
        Traders are randomly distributed across the four data centers unless otherwise noted.
        Every configuration of ABMMS has an observer, located at the Carteret node, that exports data from the simulation for analysis.
    }
    \label{fig:abmms_config}
\end{figure*}

Figure~\ref{fig:abmms_config} shows the default configuration for ABMMS, which is derived from the state of the NMS in early 2021 and adopts the propagation delays presented by ~\citet{tivnan2020fragmentation}.
ABMMS uses a discrete-event scheduler to process messages passed between agents via the CN, capturing the temporal heterogeneity of market events and agent response times.
Messages are processed sequentially based on the time they should arrive at their recipient, resulting in a dynamic step size for the simulation clock.
Exchanges are distributed across the nodes of the CN, where each exchange manages a CDA for each actively traded stock.
All CDAs in ABMMS prioritize the execution of orders based on price, visibility, and time, with ties broken randomly.
Each Exchange implements a fee schedule that includes market access fees, also known as maker-taker fees, which incentivize liquidity demand or supply depending on the configuration.
The default configuration of ABMMS includes a pair of SIPs that construct and disseminate NBBOs, LULD bands, and TAQ feeds.
See Appendix~\ref{sec:odd} for an in depth description of ABMMS following the Overview, Design concepts, Details (ODD) protocol~\cite{grimm2006standard, grimm2010odd, grimm2020odd}.

When compared with previous ABFMs, ABMMS implements several market elements that are usually abstracted away, and have never been investigated simultaneously in a single model.
Specifically, CDAs to facilitate trading, multiple assets traded simultaneously, market fragmentation in excess of two exchanges, SIPs that issue NBBOs as well as LULD bands, trade-through protection, common order modifiers (hidden, immediate-or-cancel, all-or-nothing, inter-market sweep), and market access fees.
There is an expectation of emergent phenomena in ABFMs, thus the inclusion of these additional details may have non-trivial impacts on market dynamics, especially if leveraged strategically by a learning agent.

\subsection{Traders}\label{subsec:traders}
We developed our adaptive trading agent using meta-reinforcement learning~\cite{duan2016rl, wang2016learning}.
Meta-reinforcement learning is better able to adapt to dynamic environments than traditional reinforcement learning, and financial markets are extraordinarily dynamic.
One mechanism that causes meta-reinforcement learning to foster adaptability is the use of effective experimentation processes.
Meta-reinforcement learning agents are able to actively investigate the state of their environment and incorporate that information into their decision process~\cite{dasgupta2019causal}.

Given the importance of agent adaptability and heterogeneity discussed earlier, one approach to developing reinforcement learning traders might populate a simulation entirely with such reinforcement learning traders in order to develop a population of co-adapted strategies.
However, multi-agent reinforcement learning is unstable~\cite{bucsoniu2010multi}.
With each agent adapting in real time, the optimal strategy for all agents becomes a moving goal that is difficult to approach.
Instead, we focus on the impact of a single reinforcement learning agent in simulations otherwise populated with simple agents.
For this purpose we select ZIP agents, since they have a long history of effective applications in ABFMs and have not been clearly bested by another simple strategy~\cite{rollins2020trading}.

We develop our ZIP traders based on the reference implementation provided by~\citet{cliff2018bse}, with one minor deviation.
The original implementation of ZIP traders uses an exogenous stream of limit prices as a basis for the pricing beliefs of each agent.
We replace this exogenous input with random limit prices drawn from a truncated normal distribution that is parameterized based on the current Limit Up-Limit Down (LULD) bands and is updated each time new LULD bands are issued (details in Appendix~\ref{subsubsec:zip_limit_price}).

\subsection{Stylized Facts}\label{subsec:stylized-facts}
The econometrics community has been developing stylized facts that capture various features of data generated by financial markets since the mid 90's, if not earlier~\cite{pagan1996econometrics, cont2001empirical, bouchaud2002statistical, potters2003more, sewell2011characterization}.
Stylized facts are statistical properties that are observed across a broad range of assets, markets, and time periods.
Stylized facts are qualitative and trade off precision in favor of generality, thus there can be exceptions.
However, through the combination of many stylized facts, it becomes possible to identify data that has been generated by authentic trading processes.

We focus on the eleven stylized facts outlined by~\citet{cont2001empirical}, since they are relatively simple to test for with moderate amounts of data.
However, three facts (\#1: Absence of Linear Auto-correlation, \#4: Aggregational Gaussianity, and \#11: Asymmetry in Time Scales) require long periods of coarse grained data that can be costly to generate with high fidelity models.
We eschew the three problematic facts, and rely on the remaining eight to validate ABMMS\@.

The stylized facts described by ~\citet{cont2001empirical} are exclusively concerned with properties of asset price time series.
However, ABMMS produces much more information than asset price time series.
In particular, we have access to a complete depth-of-book feed, thus metrics that investigate limit order book properties~\cite{bouchaud2002statistical, potters2003more, sewell2011characterization, paddrik2017effects} could also help to quantify the impacts of our meta-reinforcement learning trader.

\section{Results}\label{sec:abmms-results}
To ensure that our tests for stylized facts are effective, we calibrate them on historical price data.
We source minute resolution data from Alpha Vantage~\cite{av2021homepage} for 30 US stocks: AAPL, AXP, BA, CAT, CSCO, CVX, DD, DIS, GE, GS, HD, IBM, INTC, JNJ, JPM, KO, MCD, MMM, MRK, MSFT, NKE, PFE, PG, RTX, TRV, UNH, V, VZ, WMT, and XOM\@.
The data for most symbols covers two years of trading, roughly from April 2019 through February 2021.
RTX was formed as the result of a merger in 2020, and thus has truncated data coverage from April 2020 through May 2021 (roughly 14 months).
Alpha Vantage derives their data from SIP feeds and aggregates to minute resolution, with open, high, low, close, and volume features.
The data is adjusted to account for splits and dividends.
See the Alpha Vantage documentation for more details~\cite{av2021docs}.

\begin{table*}
    \centering
    \caption{
        Summarized results from the calibration of stylized fact tests.
        Stylized facts that were not confirmed in more than half of the stocks after calibration were not considered for evaluating the ABFM.
    }
    \begin{tabularx}{\textwidth}{l>{\raggedright\arraybackslash}Xll}
        Stylized Fact & Free Parameters & Best Parameter Values & Pass Rate \\
        \hline
        \#2: Heavy Tailed Returns & Window Size & max(len(returns) // 1000, 30) & 29 / 30 \\
        \#3: Asymmetry of Returns  & Window Size & max(len(returns) // 1000, 390) & 10 / 30 \\
        \#5: Intermittency of Returns & Window Size & max(len(returns) // 60, 100) & 29 / 30 \\
        \#6: Volatility Clustering & Lag Count & 5000 & 30 / 30 \\
        \#7: Heavy Tailed Conditional Returns & Window Size & max(len(returns) // 1000, 30) & 29 / 30 \\
        \#8: Slow Decay of Return Autocorrelation & Lag Count & 100 or 10000 & 21 / 30 \\
        \#9: Leverage Effect & \mbox{R Value Threshold}, \mbox{P Value Threshold} & None, None & 25 / 30 \\
        \#10: Volume/Volatility Correlation & \mbox{R Value Threshold}, \mbox{P Value Threshold} & None, None & 12 / 30 \\
    \end{tabularx}
    \label{tab:fact_calibration}
\end{table*}

We calibrate our tests for stylized facts by optimizing free parameters to improve their detection rate.
This relies on the assumption that the selected stylized facts should be expected in these stocks and during this time period.
However, stylized facts are not without exceptions and market dynamics may have qualitatively changed since the early 2000's.
Table~\ref{tab:fact_calibration} summarizes the results of our stylized fact calibration, with the main take-away being that facts \#3 and \#10 were difficult to reliably detect.
Due to this lack of consistency, we rely on the remaining six stylized facts (\#2 and \#5 through \#9) when validating ABMMS\@.

To provide the appropriate context for interpreting the impact of a learning agent, we select three control configurations.
In \texttt{zip\_simple}, we use a single exchange, a single SIP, and 30 ZIP traders, all of which are located at the Carteret node of the CN\@.
The \texttt{zip\_nms} configuration features a relatively complete representation of the NMS, with 16 exchanges distributed across the four nodes of the CN, a SIP located in Mahwah, and a SIP located in Carteret.
This condition is populated with 29 ZIP traders that are randomly distributed and one Arbitrage trader located at Secaucus.
The final condition, \texttt{zip\_no\_arb\_nms}, is identical to \texttt{zip\_nms} except that it replaces the Arbitrage trader with a ZIP trader.
Between these three configurations we can isolate the impacts of market infrastructure differences and understand some of the effects of market fragmentation.
The experimental condition is identical to \texttt{zip\_nms}, but replaces the Arbitrage trader with a Reinforcement Learning trader.

\begin{figure*}
    \centering
    \includegraphics[width=0.49\textwidth]{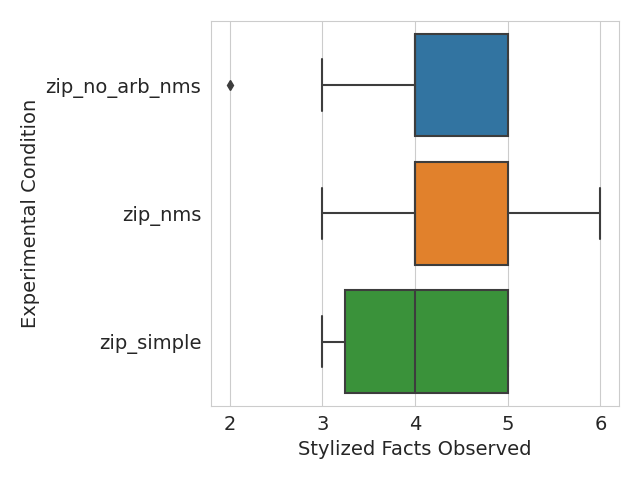}
    \caption{
        Box and whisker plots summarizing the number of stylized facts detected for each experimental condition.
        The three experimental conditions display similar capabilities for reproducing stylized facts, with an average (standard deviation) of 4.26 (0.8), 4.1 (0.78), and 4.03 (0.76) for the \texttt{zip\_no\_arb\_nms}, \texttt{zip\_nms}, and \texttt{zip\_simple} conditions respectively.
        These differences were not statistically significant according to a two sided t-test.
        \texttt{zip\_nms} was the only condition able to display all six stylized facts simultaneously.
    }
    \label{fig:cont_facts_conditions}
\end{figure*}

We collect data from 30 independent trials for each condition, where a single trial covers five trading days.
Figure~\ref{fig:cont_facts_conditions} shows the ability of data generated by each condition to display the six stylized facts that were selected based on the calibration discussed above.
All of the baseline configurations display roughly four of the six stylized facts.
The two conditions with NMS-inspired market infrastructure, \texttt{zip\_nms} and \texttt{zip\_no\_arb\_nms}, had a slight advantage over \texttt{zip\_simple}, but that difference was not statistically significant.

\begin{figure*}
    \centering
    \includegraphics[width=0.49\textwidth]{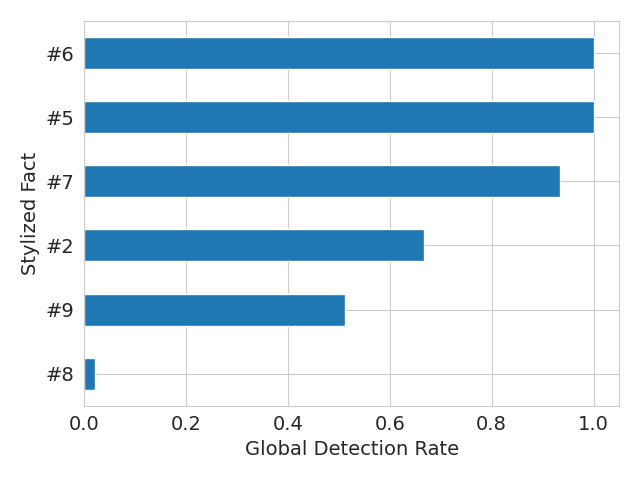}
    \includegraphics[width=0.49\textwidth]{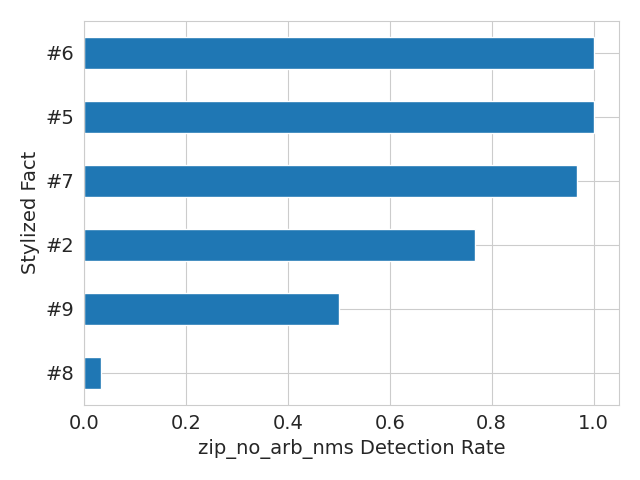}
    \includegraphics[width=0.49\textwidth]{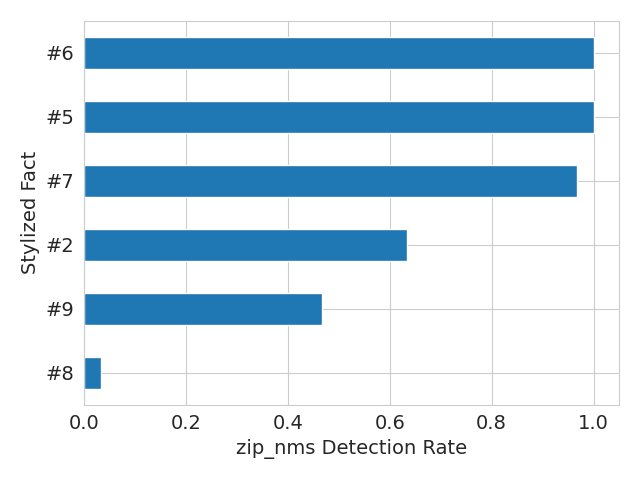}
    \includegraphics[width=0.49\textwidth]{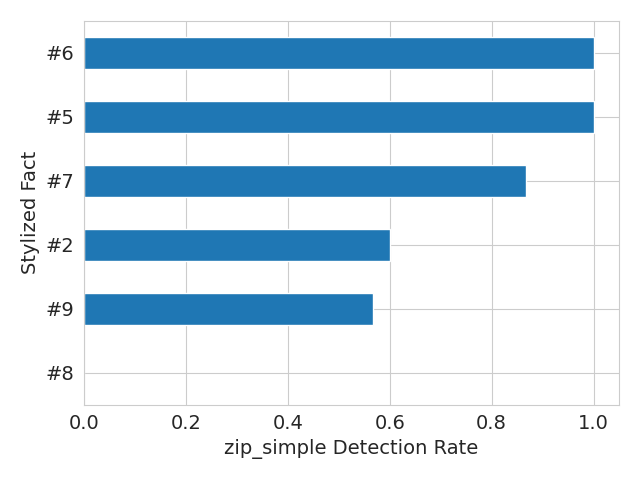}
    \caption{
        The detection rate for each stylized fact across all trials (top-left), \texttt{zip\_no\_arb\_nms} trials (top-right), \texttt{zip\_nms} trials (bottom-left), and \texttt{zip\_simple} trials (bottom-right).
        The \texttt{zip\_simple} condition was unable to display fact \#8, while both \texttt{zip\_nms} and \texttt{zip\_no\_arb\_nms} were able to display fact \#8 exactly once.
    }
    \label{fig:cont_facts_breakdown}
\end{figure*}

Figure~\ref{fig:cont_facts_breakdown} displays the detection rate of each stylized fact, across all trials and by experimental condition.
Facts \#5 and \#6 had a perfect detection rate, facts \#2, \#7, and \#9 were detected in more than 50\% of trials, and fact \#8 was detected in less than 10\% of trials.
The two conditions with NMS-inspired infrastructure were more likely to display fact \#2 and less likely to display fact \#9 than the condition with simple infrastructure.
Additionally, the condition with simple infrastructure was unable to produce a single trial that displayed fact \#8, whereas the NMS-inspired conditions both produced a single trial that did.

\begin{figure*}
    \centering
    \includegraphics[width=0.49\textwidth]{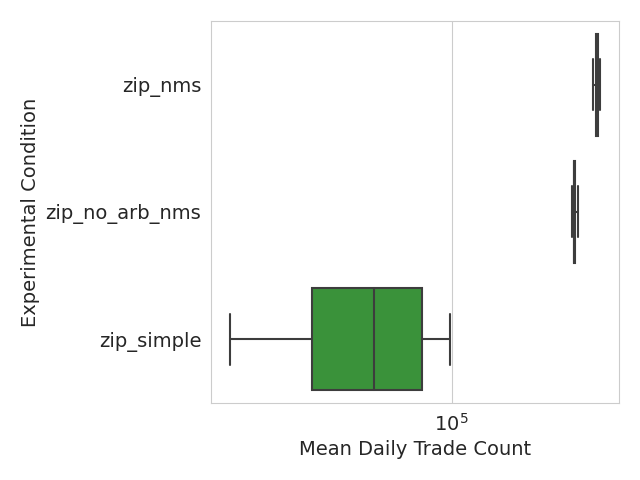}
    \includegraphics[width=0.49\textwidth]{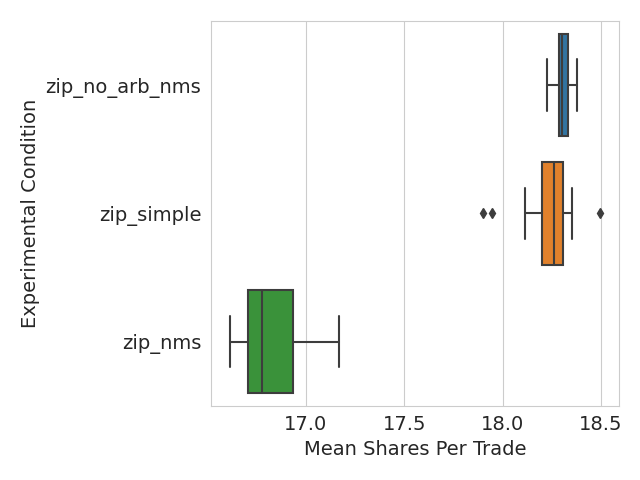}
    \includegraphics[width=0.49\textwidth]{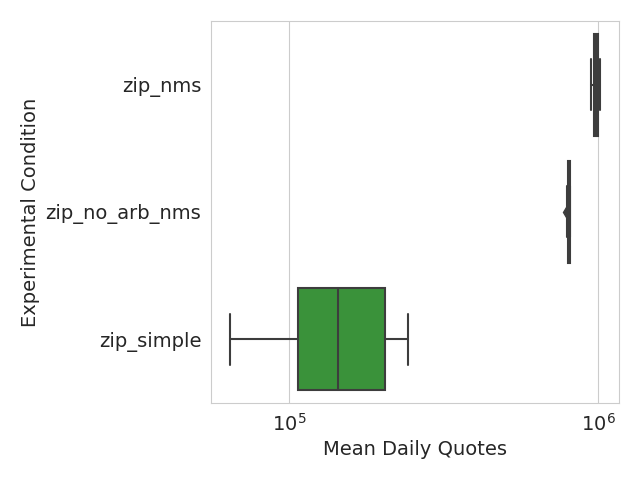}
    \includegraphics[width=0.49\textwidth]{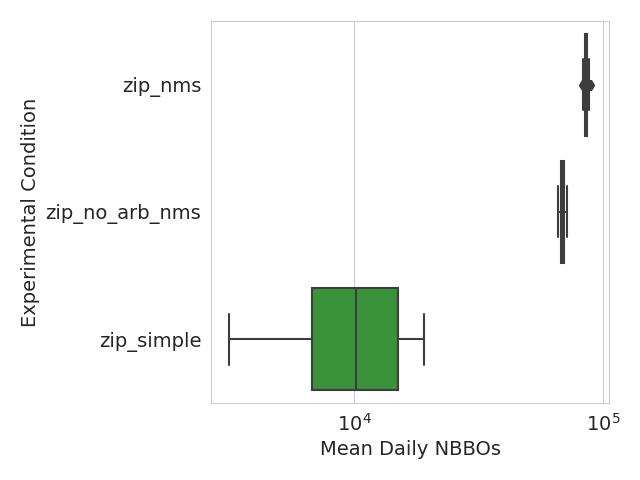}
    \caption{
        Basic trading day statistics for each experimental condition.
        The arbitrage trader causes a noticeable drop in the mean number of shares per trade (top-left).
        Market fragmentation leads to an order of magnitude increase in trades (top-right), quotes (bottom-left), and NBBOs (bottom-right).
        The arbitrage trader leads to a sizeable increase in trades, quotes, and NBBOs, but has a smaller impact than market fragmentation.
    }
    \label{fig:trade_stats_summary}
\end{figure*}

\begin{figure*}
    \centering
    \includegraphics[width=0.49\textwidth]{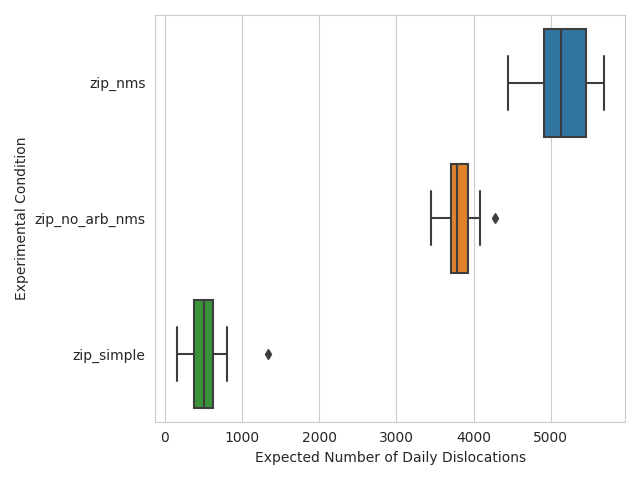}
    \includegraphics[width=0.49\textwidth]{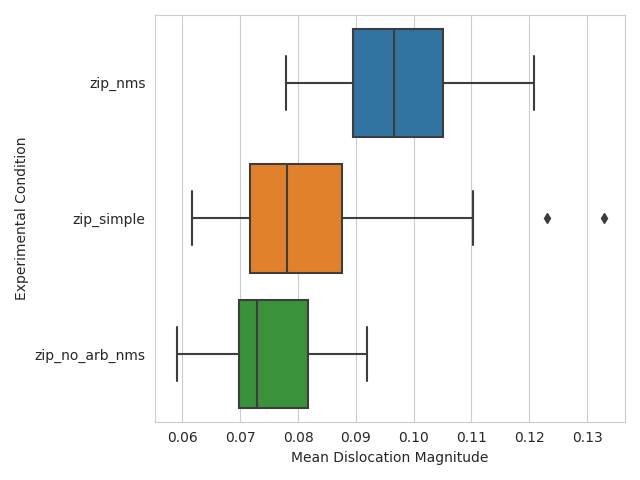}
    \includegraphics[width=0.49\textwidth]{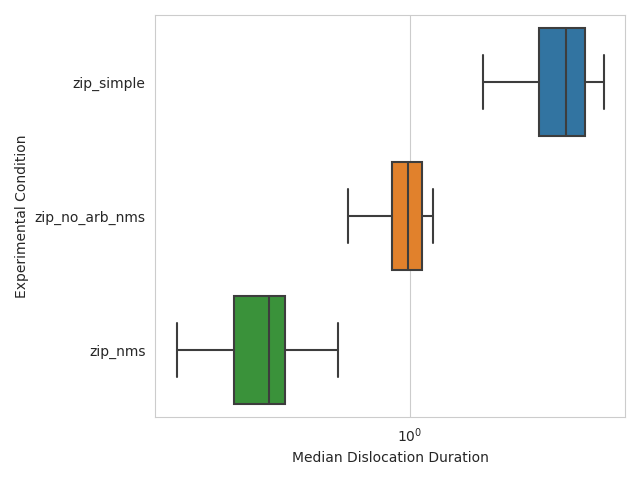}
    \caption{
        Summary statistics for dislocations by experimental condition.
        Fragmented configurations of ABMMS display roughly five times as many dislocations when compared with \texttt{zip\_simple} (top-left).
        The arbitrage trader leads to an increase in the number of dislocations (top-left) and an increase in the mean dislocation magnitude (top-right), but a decrease in the dislocation duration (bottom).
    }
    \label{fig:dislocations_summary}
\end{figure*}

The stylized facts developed by \citet{cont2001empirical} are exclusively concerned with properties of asset price time series.
However, there is a wealth of additional information that is produced by real markets, and by ABMMS\@.
Figure~\ref{fig:trade_stats_summary} investigates differences between the experimental conditions using daily occurrences of trades, quotes, and NBBOs.
Market fragmentation and the arbitrage trader both have non-trivial impacts on all of these statistics, but market fragmentation has a much larger effect.
Figure~\ref{fig:dislocations_summary} summarizes the occurrence of dislocations, as discussed in \citet{tivnan2020fragmentation}, in ABMMS\@.
The \texttt{zip\_simple} condition generates roughly an order of magnitude less dislocations than the conditions with NMS-inspired infrastructure, but those dislocations tend to be longer.
The arbitrage trader appears to cause an increase in the mean dislocation magnitude, but also a large decrease in dislocation duration.

% \todo{CVO: Describe the impacts of the RL agent in the experimental condition.}

\section{Discussion and Conclusion}\label{sec:abmms-conclusion}
% \todo{CVO: Discuss the results and their implications.}

The fact that dislocations occur in ABMMS, and can be directly measured in the same way as the NMS, is an advance in ABFM infrastructure.
\citet{tivnan2020fragmentation} indicates that dislocations in the NMS tend to have a median duration between $10^{-4}$ and $10^{-2}$ seconds, which is roughly 3 orders of magnitude smaller than the median duration displayed by the different configurations of ABMMS\@.
This is likely caused by the relatively small number of trading agents simulated, and may also indicate that we need representative strategies that characteristically operate at higher frequencies.

There are three major directions that this work could be extended.
First, we investigated the impacts of a single learning agent to avoid development difficulties that can be encountered when multiple learning agents interact, however future work should tackle these issues and develop populations of heterogeneous learning agents.
Second, we captured many important mechanisms in our implementation of ABMMS, but the NMS is an extremely complicated system and there are bound to be details that we have abstracted away.  
Enumerating and implementing these additional mechanisms will improve the accuracy of future models, and open additional strategies for learning agents to explore.
Third, we chose to exclusively implement an equities market.
However, real equity markets are linked with several financial systems, including lending systems and options markets.
Extending ABMMS to account for any of these additional financial systems could enrich the produced results.
Beyond these direct extensions, our implementation and calibration of tests for stylized facts indicates a need to revisit some common stylized facts, which may be more difficult to identify, or may not be displayed in the same ways as previously observed.

Financial market policy has been shaped largely by public comments~\cite{sec2021public}, recent events~\cite{sec2010flashcrash}, and live pilots~\cite{sec2021tickpilot}.
However, each of these influences is problematic in its own way.
Public comments can be subjective or self serving, recent events only help retrospectively, and live pilots impose costs on exchanges~\cite{michaels2020pilot}.
There have been a few successful applications of ABFMs to policy evaluation~\cite{soramaki2006topology, darley2007nasdaq, haldane2011systemic, laine2015quantitative, bookstaber2016toward, sec2017abm, bookstaber2018agent}, but additional efforts could increase the amount of policy informed by ABFMs and avoid the noted issues associated with other policy influencing mechanisms.

We do not openly provide the source code for our agents, models, and analysis due to concerns about potential malicious use.
We will consider sharing all or part of the source code on a case by case basis.
Please contact the authors if interested.
Finally, to facilitate peer review and some degree of transparency, we provide all of the model outputs from experimental runs used to create the figures and analysis of this paper~\cite{vanoort2021abmms}.

\section*{Acknowledgements}
We thank Thayer Alshaabi, David Dewhurst, Matthew Koehler, and John Ring for their insightful discussion and suggestions.
Some computations were performed on the Vermont Advanced Computing Core, supported in part by NSF award No.~OAC-1827314.

\bibliographystyle{ACM-Reference-Format}
\bibliography{references}

%%% -*-BibTeX-*-
%%% Do NOT edit. File created by BibTeX with style
%%% ACM-Reference-Format-Journals [18-Jan-2012].

\begin{thebibliography}{95}

%%% ====================================================================
%%% NOTE TO THE USER: you can override these defaults by providing
%%% customized versions of any of these macros before the \bibliography
%%% command.  Each of them MUST provide its own final punctuation,
%%% except for \shownote{}, \showDOI{}, and \showURL{}.  The latter two
%%% do not use final punctuation, in order to avoid confusing it with
%%% the Web address.
%%%
%%% To suppress output of a particular field, define its macro to expand
%%% to an empty string, or better, \unskip, like this:
%%%
%%% \newcommand{\showDOI}[1]{\unskip}   % LaTeX syntax
%%%
%%% \def \showDOI #1{\unskip}           % plain TeX syntax
%%%
%%% ====================================================================

\ifx \showCODEN    \undefined \def \showCODEN     #1{\unskip}     \fi
\ifx \showDOI      \undefined \def \showDOI       #1{#1}\fi
\ifx \showISBNx    \undefined \def \showISBNx     #1{\unskip}     \fi
\ifx \showISBNxiii \undefined \def \showISBNxiii  #1{\unskip}     \fi
\ifx \showISSN     \undefined \def \showISSN      #1{\unskip}     \fi
\ifx \showLCCN     \undefined \def \showLCCN      #1{\unskip}     \fi
\ifx \shownote     \undefined \def \shownote      #1{#1}          \fi
\ifx \showarticletitle \undefined \def \showarticletitle #1{#1}   \fi
\ifx \showURL      \undefined \def \showURL       {\relax}        \fi
% The following commands are used for tagged output and should be
% invisible to TeX
\providecommand\bibfield[2]{#2}
\providecommand\bibinfo[2]{#2}
\providecommand\natexlab[1]{#1}
\providecommand\showeprint[2][]{arXiv:#2}

\bibitem[\protect\citeauthoryear{Abergel, Anane, Chakraborti, Jedidi, and
  Toke}{Abergel et~al\mbox{.}}{2016}]%
        {abergel2016limit}
\bibfield{author}{\bibinfo{person}{Fr{\'e}d{\'e}ric Abergel},
  \bibinfo{person}{Marouane Anane}, \bibinfo{person}{Anirban Chakraborti},
  \bibinfo{person}{Aymen Jedidi}, {and} \bibinfo{person}{Ioane~Muni Toke}.}
  \bibinfo{year}{2016}\natexlab{}.
\newblock \bibinfo{booktitle}{\emph{Limit order books}}.
\newblock \bibinfo{publisher}{Cambridge University Press}.
\newblock


\bibitem[\protect\citeauthoryear{Arifin and Madey}{Arifin and Madey}{2015}]%
        {arifin2015verification}
\bibfield{author}{\bibinfo{person}{SM~Niaz Arifin} {and}
  \bibinfo{person}{Gregory~R Madey}.} \bibinfo{year}{2015}\natexlab{}.
\newblock \showarticletitle{Verification, Validation, and Replication Methods
  for Agent-Based Modeling and Simulation: Lessons Learned the Hard Way!}
\newblock In \bibinfo{booktitle}{\emph{Concepts and Methodologies for Modeling
  and Simulation}}. \bibinfo{publisher}{Springer}, \bibinfo{pages}{217--242}.
\newblock


\bibitem[\protect\citeauthoryear{Baldauf and Mollner}{Baldauf and
  Mollner}{2021}]%
        {baldauf2021fragmented}
\bibfield{author}{\bibinfo{person}{Markus Baldauf} {and}
  \bibinfo{person}{Joshua Mollner}.} \bibinfo{year}{2021}\natexlab{}.
\newblock \showarticletitle{Trading in fragmented markets}.
\newblock \bibinfo{journal}{\emph{Journal of Financial and Quantitative
  Analysis}} \bibinfo{volume}{56}, \bibinfo{number}{1} (\bibinfo{date}{2}
  \bibinfo{year}{2021}), \bibinfo{pages}{93--121}.
\newblock


\bibitem[\protect\citeauthoryear{Bookstaber, Foley, and Tivnan}{Bookstaber
  et~al\mbox{.}}{2016}]%
        {bookstaber2016toward}
\bibfield{author}{\bibinfo{person}{Richard Bookstaber},
  \bibinfo{person}{Michael~D Foley}, {and} \bibinfo{person}{Brian~F Tivnan}.}
  \bibinfo{year}{2016}\natexlab{}.
\newblock \showarticletitle{Toward an understanding of market resilience:
  market liquidity and heterogeneity in the investor decision cycle}.
\newblock \bibinfo{journal}{\emph{Journal of Economic Interaction and
  Coordination}} \bibinfo{volume}{11}, \bibinfo{number}{2}
  (\bibinfo{year}{2016}), \bibinfo{pages}{205--227}.
\newblock


\bibitem[\protect\citeauthoryear{Bookstaber, Paddrik, and Tivnan}{Bookstaber
  et~al\mbox{.}}{2018}]%
        {bookstaber2018agent}
\bibfield{author}{\bibinfo{person}{Richard Bookstaber}, \bibinfo{person}{Mark
  Paddrik}, {and} \bibinfo{person}{Brian Tivnan}.}
  \bibinfo{year}{2018}\natexlab{}.
\newblock \showarticletitle{An agent-based model for financial vulnerability}.
\newblock \bibinfo{journal}{\emph{Journal of Economic Interaction and
  Coordination}} \bibinfo{volume}{13}, \bibinfo{number}{2}
  (\bibinfo{year}{2018}), \bibinfo{pages}{433--466}.
\newblock


\bibitem[\protect\citeauthoryear{Bouchaud, M{\'e}zard, and Potters}{Bouchaud
  et~al\mbox{.}}{2002}]%
        {bouchaud2002statistical}
\bibfield{author}{\bibinfo{person}{Jean-Philippe Bouchaud},
  \bibinfo{person}{Marc M{\'e}zard}, {and} \bibinfo{person}{Marc Potters}.}
  \bibinfo{year}{2002}\natexlab{}.
\newblock \showarticletitle{Statistical properties of stock order books:
  empirical results and models}.
\newblock \bibinfo{journal}{\emph{Quantitative finance}}  \bibinfo{volume}{2}
  (\bibinfo{year}{2002}), \bibinfo{pages}{251--256}.
\newblock


\bibitem[\protect\citeauthoryear{Brown}{Brown}{2011}]%
        {brown2011efficient}
\bibfield{author}{\bibinfo{person}{Stephen~J Brown}.}
  \bibinfo{year}{2011}\natexlab{}.
\newblock \showarticletitle{The efficient markets hypothesis: The demise of the
  demon of chance?}
\newblock \bibinfo{journal}{\emph{Accounting \& Finance}} \bibinfo{volume}{51},
  \bibinfo{number}{1} (\bibinfo{year}{2011}), \bibinfo{pages}{79--95}.
\newblock


\bibitem[\protect\citeauthoryear{Bu{\c{s}}oniu, Babu{\v{s}}ka, and
  De~Schutter}{Bu{\c{s}}oniu et~al\mbox{.}}{2010}]%
        {bucsoniu2010multi}
\bibfield{author}{\bibinfo{person}{Lucian Bu{\c{s}}oniu},
  \bibinfo{person}{Robert Babu{\v{s}}ka}, {and} \bibinfo{person}{Bart
  De~Schutter}.} \bibinfo{year}{2010}\natexlab{}.
\newblock \showarticletitle{Multi-agent reinforcement learning: An overview}.
\newblock \bibinfo{journal}{\emph{Innovations in multi-agent systems and
  applications-1}} (\bibinfo{year}{2010}), \bibinfo{pages}{183--221}.
\newblock


\bibitem[\protect\citeauthoryear{Byrd, Hybinette, and Balch}{Byrd
  et~al\mbox{.}}{2019}]%
        {byrd2019abides}
\bibfield{author}{\bibinfo{person}{David Byrd}, \bibinfo{person}{Maria
  Hybinette}, {and} \bibinfo{person}{Tucker~Hybinette Balch}.}
  \bibinfo{year}{2019}\natexlab{}.
\newblock \showarticletitle{Abides: Towards high-fidelity market simulation for
  ai research}.
\newblock \bibinfo{journal}{\emph{arXiv preprint arXiv:1904.12066}}
  (\bibinfo{year}{2019}).
\newblock


\bibitem[\protect\citeauthoryear{Cliff}{Cliff}{1997}]%
        {cliff1997minimal}
\bibfield{author}{\bibinfo{person}{Dave Cliff}.}
  \bibinfo{year}{1997}\natexlab{}.
\newblock \showarticletitle{Minimal-intelligence agents for bargaining
  behaviors in market-based environments}.
\newblock \bibinfo{journal}{\emph{Hewlett-Packard Labs Technical Reports}}
  (\bibinfo{year}{1997}).
\newblock


\bibitem[\protect\citeauthoryear{Cliff}{Cliff}{2018}]%
        {cliff2018bse}
\bibfield{author}{\bibinfo{person}{Dave Cliff}.}
  \bibinfo{year}{2018}\natexlab{}.
\newblock \showarticletitle{BSE: A Minimal Simulation of a Limit-Order-Book
  Stock Exchange}.
\newblock \bibinfo{journal}{\emph{Proceedings of the 30th European Modeling and
  Simulation Symposium (EMSS 2018)}} (\bibinfo{year}{2018}),
  \bibinfo{pages}{194--203}.
\newblock


\bibitem[\protect\citeauthoryear{Cliff and Bruten}{Cliff and Bruten}{1997}]%
        {cliff1997zero}
\bibfield{author}{\bibinfo{person}{Dave Cliff} {and} \bibinfo{person}{Janet
  Bruten}.} \bibinfo{year}{1997}\natexlab{}.
\newblock \showarticletitle{Zero is Not Enough: On The Lower Limit of Agent
  Intelligence For Continuous Double Auction Markets}.
\newblock \bibinfo{journal}{\emph{Hewlett-Packard Labs Technical Reports}}
  (\bibinfo{year}{1997}).
\newblock


\bibitem[\protect\citeauthoryear{Collver~Jr.}{Collver~Jr.}{2017}]%
        {sec2017abm}
\bibfield{author}{\bibinfo{person}{Charles~D. Collver~Jr.}}
  \bibinfo{year}{2017}\natexlab{}.
\newblock \bibinfo{title}{An application of agent-based modeling to market
  structure policy: the case of the U.S. Tick Size Pilot Program and market
  maker profitability}.
\newblock
\newblock
\urldef\tempurl%
\url{https://www.sec.gov/marketstructure/research/increasing-the-mpi-combined.pdf}
\showURL{%
\tempurl}


\bibitem[\protect\citeauthoryear{Cont}{Cont}{2001}]%
        {cont2001empirical}
\bibfield{author}{\bibinfo{person}{Rama Cont}.}
  \bibinfo{year}{2001}\natexlab{}.
\newblock \showarticletitle{Empirical properties of asset returns: stylized
  facts and statistical issues}.
\newblock \bibinfo{journal}{\emph{Quantitative Finance}} \bibinfo{volume}{1},
  \bibinfo{number}{1} (\bibinfo{year}{2001}), \bibinfo{pages}{223--236}.
\newblock


\bibitem[\protect\citeauthoryear{Cont, Stoikov, and Talreja}{Cont
  et~al\mbox{.}}{2010}]%
        {cont2010stochastic}
\bibfield{author}{\bibinfo{person}{Rama Cont}, \bibinfo{person}{Sasha Stoikov},
  {and} \bibinfo{person}{Rishi Talreja}.} \bibinfo{year}{2010}\natexlab{}.
\newblock \showarticletitle{A stochastic model for order book dynamics}.
\newblock \bibinfo{journal}{\emph{Operations research}} \bibinfo{volume}{58},
  \bibinfo{number}{3} (\bibinfo{year}{2010}), \bibinfo{pages}{549--563}.
\newblock


\bibitem[\protect\citeauthoryear{Darley}{Darley}{2007}]%
        {darley2007nasdaq}
\bibfield{author}{\bibinfo{person}{Vincent Darley}.}
  \bibinfo{year}{2007}\natexlab{}.
\newblock \bibinfo{booktitle}{\emph{A NASDAQ market simulation: insights on a
  major market from the science of complex adaptive systems}}.
  Vol.~\bibinfo{volume}{1}.
\newblock \bibinfo{publisher}{World Scientific}.
\newblock


\bibitem[\protect\citeauthoryear{Dasgupta, Wang, Chiappa, Mitrovic, Ortega,
  Raposo, Hughes, Battaglia, Botvinick, and Kurth-Nelson}{Dasgupta
  et~al\mbox{.}}{2019}]%
        {dasgupta2019causal}
\bibfield{author}{\bibinfo{person}{Ishita Dasgupta}, \bibinfo{person}{Jane
  Wang}, \bibinfo{person}{Silvia Chiappa}, \bibinfo{person}{Jovana Mitrovic},
  \bibinfo{person}{Pedro Ortega}, \bibinfo{person}{David Raposo},
  \bibinfo{person}{Edward Hughes}, \bibinfo{person}{Peter Battaglia},
  \bibinfo{person}{Matthew Botvinick}, {and} \bibinfo{person}{Zeb
  Kurth-Nelson}.} \bibinfo{year}{2019}\natexlab{}.
\newblock \showarticletitle{Causal reasoning from meta-reinforcement learning}.
\newblock \bibinfo{journal}{\emph{arXiv preprint arXiv:1901.08162}}
  (\bibinfo{year}{2019}).
\newblock


\bibitem[\protect\citeauthoryear{Deng, Bao, Kong, Ren, and Dai}{Deng
  et~al\mbox{.}}{2016}]%
        {deng2016deep}
\bibfield{author}{\bibinfo{person}{Yue Deng}, \bibinfo{person}{Feng Bao},
  \bibinfo{person}{Youyong Kong}, \bibinfo{person}{Zhiquan Ren}, {and}
  \bibinfo{person}{Qionghai Dai}.} \bibinfo{year}{2016}\natexlab{}.
\newblock \showarticletitle{Deep direct reinforcement learning for financial
  signal representation and trading}.
\newblock \bibinfo{journal}{\emph{IEEE transactions on neural networks and
  learning systems}} \bibinfo{volume}{28}, \bibinfo{number}{3}
  (\bibinfo{year}{2016}), \bibinfo{pages}{653--664}.
\newblock


\bibitem[\protect\citeauthoryear{Dewhurst, Van~Oort, Ring~IV, Gray, Danforth,
  and Tivnan}{Dewhurst et~al\mbox{.}}{2019}]%
        {dewhurst2019scaling}
\bibfield{author}{\bibinfo{person}{David~Rushing Dewhurst},
  \bibinfo{person}{Colin~M Van~Oort}, \bibinfo{person}{John~H Ring~IV},
  \bibinfo{person}{Tyler~J Gray}, \bibinfo{person}{Christopher~M Danforth},
  {and} \bibinfo{person}{Brian~F Tivnan}.} \bibinfo{year}{2019}\natexlab{}.
\newblock \showarticletitle{Scaling of inefficiencies in the US equity markets:
  Evidence from three market indices and more than 2900 securities}.
\newblock \bibinfo{journal}{\emph{arXiv preprint arXiv:1902.04691}}
  (\bibinfo{year}{2019}).
\newblock


\bibitem[\protect\citeauthoryear{Di~Mascio, Lines, and Naik}{Di~Mascio
  et~al\mbox{.}}{2016}]%
        {di2016alpha}
\bibfield{author}{\bibinfo{person}{Rick Di~Mascio}, \bibinfo{person}{Anton
  Lines}, {and} \bibinfo{person}{Narayan~Y Naik}.}
  \bibinfo{year}{2016}\natexlab{}.
\newblock \showarticletitle{Alpha decay}.
\newblock \bibinfo{journal}{\emph{SFS Finance Cavalcade}}
  (\bibinfo{year}{2016}).
\newblock


\bibitem[\protect\citeauthoryear{Dicks and Gebbie}{Dicks and Gebbie}{2022}]%
        {dicks2022simple}
\bibfield{author}{\bibinfo{person}{Matthew Dicks} {and} \bibinfo{person}{Tim
  Gebbie}.} \bibinfo{year}{2022}\natexlab{}.
\newblock \showarticletitle{A simple learning agent interacting with an
  agent-based market model}.
\newblock \bibinfo{journal}{\emph{arXiv preprint arXiv:2208.10434}}
  (\bibinfo{year}{2022}).
\newblock


\bibitem[\protect\citeauthoryear{Duan, Schulman, Chen, Bartlett, Sutskever, and
  Abbeel}{Duan et~al\mbox{.}}{2016}]%
        {duan2016rl}
\bibfield{author}{\bibinfo{person}{Yan Duan}, \bibinfo{person}{John Schulman},
  \bibinfo{person}{Xi Chen}, \bibinfo{person}{Peter~L Bartlett},
  \bibinfo{person}{Ilya Sutskever}, {and} \bibinfo{person}{Pieter Abbeel}.}
  \bibinfo{year}{2016}\natexlab{}.
\newblock \showarticletitle{Rl $^2$: Fast reinforcement learning via slow
  reinforcement learning}.
\newblock \bibinfo{journal}{\emph{arXiv preprint arXiv:1611.02779}}
  (\bibinfo{year}{2016}).
\newblock


\bibitem[\protect\citeauthoryear{Duffin and Cartlidge}{Duffin and
  Cartlidge}{2018}]%
        {duffin2018agent}
\bibfield{author}{\bibinfo{person}{Matthew Duffin} {and} \bibinfo{person}{John
  Cartlidge}.} \bibinfo{year}{2018}\natexlab{}.
\newblock \showarticletitle{Agent-based model exploration of latency arbitrage
  in fragmented financial markets}. In \bibinfo{booktitle}{\emph{2018 IEEE
  Symposium Series on Computational Intelligence (SSCI)}}.
  \bibinfo{publisher}{IEEE}, \bibinfo{pages}{2312--2320}.
\newblock


\bibitem[\protect\citeauthoryear{Ecca, Marchesi, and Setzu}{Ecca
  et~al\mbox{.}}{2008}]%
        {ecca2008modeling}
\bibfield{author}{\bibinfo{person}{Sabrina Ecca}, \bibinfo{person}{Michele
  Marchesi}, {and} \bibinfo{person}{Alessio Setzu}.}
  \bibinfo{year}{2008}\natexlab{}.
\newblock \showarticletitle{Modeling and simulation of an artificial stock
  option market}.
\newblock \bibinfo{journal}{\emph{Computational Economics}}
  \bibinfo{volume}{32} (\bibinfo{year}{2008}), \bibinfo{pages}{37--53}.
\newblock


\bibitem[\protect\citeauthoryear{Espeholt, Soyer, Munos, Simonyan, Mnih, Ward,
  Doron, Firoiu, Harley, Dunning, et~al\mbox{.}}{Espeholt
  et~al\mbox{.}}{2018}]%
        {espeholt2018impala}
\bibfield{author}{\bibinfo{person}{Lasse Espeholt}, \bibinfo{person}{Hubert
  Soyer}, \bibinfo{person}{Remi Munos}, \bibinfo{person}{Karen Simonyan},
  \bibinfo{person}{Vlad Mnih}, \bibinfo{person}{Tom Ward},
  \bibinfo{person}{Yotam Doron}, \bibinfo{person}{Vlad Firoiu},
  \bibinfo{person}{Tim Harley}, \bibinfo{person}{Iain Dunning},
  {et~al\mbox{.}}} \bibinfo{year}{2018}\natexlab{}.
\newblock \showarticletitle{Impala: Scalable distributed deep-rl with
  importance weighted actor-learner architectures}. In
  \bibinfo{booktitle}{\emph{International Conference on Machine Learning}}.
  \bibinfo{publisher}{ICML}, \bibinfo{pages}{1407--1416}.
\newblock


\bibitem[\protect\citeauthoryear{Farmer, Patelli, and Zovko}{Farmer
  et~al\mbox{.}}{2005}]%
        {farmer2005predictive}
\bibfield{author}{\bibinfo{person}{J~Doyne Farmer}, \bibinfo{person}{Paolo
  Patelli}, {and} \bibinfo{person}{Ilija~I Zovko}.}
  \bibinfo{year}{2005}\natexlab{}.
\newblock \showarticletitle{The predictive power of zero intelligence in
  financial markets}.
\newblock \bibinfo{journal}{\emph{Proceedings of the National Academy of
  Sciences}} \bibinfo{volume}{102}, \bibinfo{number}{6} (\bibinfo{year}{2005}),
  \bibinfo{pages}{2254--2259}.
\newblock


\bibitem[\protect\citeauthoryear{Friedman}{Friedman}{2018}]%
        {friedman2018double}
\bibfield{author}{\bibinfo{person}{Daniel Friedman}.}
  \bibinfo{year}{2018}\natexlab{}.
\newblock \showarticletitle{The double auction market institution: A survey}.
\newblock In \bibinfo{booktitle}{\emph{The Double Auction Market Institutions,
  Theories, and Evidence}}. \bibinfo{publisher}{Routledge},
  \bibinfo{pages}{3--26}.
\newblock


\bibitem[\protect\citeauthoryear{Gao, Vytelingum, Weston, Luk, and Guo}{Gao
  et~al\mbox{.}}{2022a}]%
        {gao2022high}
\bibfield{author}{\bibinfo{person}{Kang Gao}, \bibinfo{person}{Perukrishnen
  Vytelingum}, \bibinfo{person}{Stephen Weston}, \bibinfo{person}{Wayne Luk},
  {and} \bibinfo{person}{Ce Guo}.} \bibinfo{year}{2022}\natexlab{a}.
\newblock \showarticletitle{High-frequency financial market simulation and
  flash crash scenarios analysis: an agent-based modelling approach}.
\newblock \bibinfo{journal}{\emph{arXiv preprint arXiv:2208.13654}}
  (\bibinfo{year}{2022}).
\newblock


\bibitem[\protect\citeauthoryear{Gao, Vytelingum, Weston, Luk, and Guo}{Gao
  et~al\mbox{.}}{2022b}]%
        {gao2022understanding}
\bibfield{author}{\bibinfo{person}{Kang Gao}, \bibinfo{person}{Perukrishnen
  Vytelingum}, \bibinfo{person}{Stephen Weston}, \bibinfo{person}{Wayne Luk},
  {and} \bibinfo{person}{Ce Guo}.} \bibinfo{year}{2022}\natexlab{b}.
\newblock \showarticletitle{Understanding intra-day price formation process by
  agent-based financial market simulation: calibrating the extended chiarella
  model}.
\newblock \bibinfo{journal}{\emph{arXiv preprint arXiv:2208.14207}}
  (\bibinfo{year}{2022}).
\newblock


\bibitem[\protect\citeauthoryear{Ghoulmie, Cont, and Nadal}{Ghoulmie
  et~al\mbox{.}}{2005}]%
        {ghoulmie2005heterogeneity}
\bibfield{author}{\bibinfo{person}{Francois Ghoulmie}, \bibinfo{person}{Rama
  Cont}, {and} \bibinfo{person}{Jean-Pierre Nadal}.}
  \bibinfo{year}{2005}\natexlab{}.
\newblock \showarticletitle{Heterogeneity and feedback in an agent-based market
  model}.
\newblock \bibinfo{journal}{\emph{Journal of Physics: condensed matter}}
  \bibinfo{volume}{17}, \bibinfo{number}{14} (\bibinfo{year}{2005}),
  \bibinfo{pages}{S1259}.
\newblock


\bibitem[\protect\citeauthoryear{Gjerstad and Dickhaut}{Gjerstad and
  Dickhaut}{1998}]%
        {gjerstad1998price}
\bibfield{author}{\bibinfo{person}{Steven Gjerstad} {and} \bibinfo{person}{John
  Dickhaut}.} \bibinfo{year}{1998}\natexlab{}.
\newblock \showarticletitle{Price formation in double auctions}.
\newblock \bibinfo{journal}{\emph{Games and economic behavior}}
  \bibinfo{volume}{22}, \bibinfo{number}{1} (\bibinfo{year}{1998}),
  \bibinfo{pages}{1--29}.
\newblock


\bibitem[\protect\citeauthoryear{Gode and Sunder}{Gode and Sunder}{1993}]%
        {gode1993allocative}
\bibfield{author}{\bibinfo{person}{Dhananjay~K Gode} {and}
  \bibinfo{person}{Shyam Sunder}.} \bibinfo{year}{1993}\natexlab{}.
\newblock \showarticletitle{Allocative efficiency of markets with
  zero-intelligence traders: Market as a partial substitute for individual
  rationality}.
\newblock \bibinfo{journal}{\emph{Journal of political economy}}
  \bibinfo{volume}{101}, \bibinfo{number}{1} (\bibinfo{year}{1993}),
  \bibinfo{pages}{119--137}.
\newblock


\bibitem[\protect\citeauthoryear{Gould, Porter, Williams, McDonald, Fenn, and
  Howison}{Gould et~al\mbox{.}}{2013}]%
        {gould2013limit}
\bibfield{author}{\bibinfo{person}{Martin~D Gould}, \bibinfo{person}{Mason~A
  Porter}, \bibinfo{person}{Stacy Williams}, \bibinfo{person}{Mark McDonald},
  \bibinfo{person}{Daniel~J Fenn}, {and} \bibinfo{person}{Sam~D Howison}.}
  \bibinfo{year}{2013}\natexlab{}.
\newblock \showarticletitle{Limit order books}.
\newblock \bibinfo{journal}{\emph{Quantitative Finance}} \bibinfo{volume}{13},
  \bibinfo{number}{11} (\bibinfo{year}{2013}), \bibinfo{pages}{1709--1742}.
\newblock


\bibitem[\protect\citeauthoryear{Grimm, Berger, Bastiansen, Eliassen, Ginot,
  Giske, Goss-Custard, Grand, Heinz, Huse, et~al\mbox{.}}{Grimm
  et~al\mbox{.}}{2006}]%
        {grimm2006standard}
\bibfield{author}{\bibinfo{person}{Volker Grimm}, \bibinfo{person}{Uta Berger},
  \bibinfo{person}{Finn Bastiansen}, \bibinfo{person}{Sigrunn Eliassen},
  \bibinfo{person}{Vincent Ginot}, \bibinfo{person}{Jarl Giske},
  \bibinfo{person}{John Goss-Custard}, \bibinfo{person}{Tamara Grand},
  \bibinfo{person}{Simone~K Heinz}, \bibinfo{person}{Geir Huse},
  {et~al\mbox{.}}} \bibinfo{year}{2006}\natexlab{}.
\newblock \showarticletitle{A standard protocol for describing individual-based
  and agent-based models}.
\newblock \bibinfo{journal}{\emph{Ecological modelling}} \bibinfo{volume}{198},
  \bibinfo{number}{1-2} (\bibinfo{year}{2006}), \bibinfo{pages}{115--126}.
\newblock


\bibitem[\protect\citeauthoryear{Grimm, Berger, DeAngelis, Polhill, Giske, and
  Railsback}{Grimm et~al\mbox{.}}{2010}]%
        {grimm2010odd}
\bibfield{author}{\bibinfo{person}{Volker Grimm}, \bibinfo{person}{Uta Berger},
  \bibinfo{person}{Donald~L DeAngelis}, \bibinfo{person}{J~Gary Polhill},
  \bibinfo{person}{Jarl Giske}, {and} \bibinfo{person}{Steven~F Railsback}.}
  \bibinfo{year}{2010}\natexlab{}.
\newblock \showarticletitle{The ODD protocol: a review and first update}.
\newblock \bibinfo{journal}{\emph{Ecological modelling}} \bibinfo{volume}{221},
  \bibinfo{number}{23} (\bibinfo{year}{2010}), \bibinfo{pages}{2760--2768}.
\newblock


\bibitem[\protect\citeauthoryear{Grimm, Railsback, Vincenot, Berger, Gallagher,
  DeAngelis, Edmonds, Ge, Giske, Groeneveld, et~al\mbox{.}}{Grimm
  et~al\mbox{.}}{2020}]%
        {grimm2020odd}
\bibfield{author}{\bibinfo{person}{Volker Grimm}, \bibinfo{person}{Steven~F
  Railsback}, \bibinfo{person}{Christian~E Vincenot}, \bibinfo{person}{Uta
  Berger}, \bibinfo{person}{Cara Gallagher}, \bibinfo{person}{Donald~L
  DeAngelis}, \bibinfo{person}{Bruce Edmonds}, \bibinfo{person}{Jiaqi Ge},
  \bibinfo{person}{Jarl Giske}, \bibinfo{person}{Juergen Groeneveld},
  {et~al\mbox{.}}} \bibinfo{year}{2020}\natexlab{}.
\newblock \showarticletitle{The ODD protocol for describing agent-based and
  other simulation models: A second update to improve clarity, replication, and
  structural realism}.
\newblock \bibinfo{journal}{\emph{Journal of Artificial Societies and Social
  Simulation}} \bibinfo{volume}{23}, \bibinfo{number}{2}
  (\bibinfo{year}{2020}).
\newblock


\bibitem[\protect\citeauthoryear{Haldane and May}{Haldane and May}{2011}]%
        {haldane2011systemic}
\bibfield{author}{\bibinfo{person}{Andrew~G Haldane} {and}
  \bibinfo{person}{Robert~M May}.} \bibinfo{year}{2011}\natexlab{}.
\newblock \showarticletitle{Systemic risk in banking ecosystems}.
\newblock \bibinfo{journal}{\emph{Nature}} \bibinfo{volume}{469},
  \bibinfo{number}{7330} (\bibinfo{year}{2011}), \bibinfo{pages}{351--355}.
\newblock


\bibitem[\protect\citeauthoryear{Hammond}{Hammond}{1997}]%
        {hammond1997rationality}
\bibfield{author}{\bibinfo{person}{Peter~J Hammond}.}
  \bibinfo{year}{1997}\natexlab{}.
\newblock \showarticletitle{Rationality in economics}.
\newblock \bibinfo{journal}{\emph{Rivista internazionale di scienze sociali}}
  \bibinfo{volume}{105}, \bibinfo{number}{3} (\bibinfo{year}{1997}),
  \bibinfo{pages}{247--288}.
\newblock


\bibitem[\protect\citeauthoryear{Hasbrouck}{Hasbrouck}{2007}]%
        {hasbrouck2007empirical}
\bibfield{author}{\bibinfo{person}{Joel Hasbrouck}.}
  \bibinfo{year}{2007}\natexlab{}.
\newblock \bibinfo{booktitle}{\emph{Empirical market microstructure: The
  institutions, economics, and econometrics of securities trading}}.
\newblock \bibinfo{publisher}{Oxford University Press}.
\newblock


\bibitem[\protect\citeauthoryear{Hochreiter and Schmidhuber}{Hochreiter and
  Schmidhuber}{1997}]%
        {hochreiter1997long}
\bibfield{author}{\bibinfo{person}{Sepp Hochreiter} {and}
  \bibinfo{person}{J{\"u}rgen Schmidhuber}.} \bibinfo{year}{1997}\natexlab{}.
\newblock \showarticletitle{Long short-term memory}.
\newblock \bibinfo{journal}{\emph{Neural computation}} \bibinfo{volume}{9},
  \bibinfo{number}{8} (\bibinfo{year}{1997}), \bibinfo{pages}{1735--1780}.
\newblock


\bibitem[\protect\citeauthoryear{Hospedales, Antoniou, Micaelli, and
  Storkey}{Hospedales et~al\mbox{.}}{2020}]%
        {hospedales2020meta}
\bibfield{author}{\bibinfo{person}{Timothy Hospedales},
  \bibinfo{person}{Antreas Antoniou}, \bibinfo{person}{Paul Micaelli}, {and}
  \bibinfo{person}{Amos Storkey}.} \bibinfo{year}{2020}\natexlab{}.
\newblock \showarticletitle{Meta-learning in neural networks: A survey}.
\newblock \bibinfo{journal}{\emph{arXiv preprint arXiv:2004.05439}}
  (\bibinfo{year}{2020}).
\newblock


\bibitem[\protect\citeauthoryear{Hu, Liu, Zhang, Su, Ngai, and Liu}{Hu
  et~al\mbox{.}}{2015}]%
        {hu2015application}
\bibfield{author}{\bibinfo{person}{Yong Hu}, \bibinfo{person}{Kang Liu},
  \bibinfo{person}{Xiangzhou Zhang}, \bibinfo{person}{Lijun Su},
  \bibinfo{person}{EWT Ngai}, {and} \bibinfo{person}{Mei Liu}.}
  \bibinfo{year}{2015}\natexlab{}.
\newblock \showarticletitle{Application of evolutionary computation for rule
  discovery in stock algorithmic trading: A literature review}.
\newblock \bibinfo{journal}{\emph{Applied Soft Computing}}
  \bibinfo{volume}{36} (\bibinfo{year}{2015}), \bibinfo{pages}{534--551}.
\newblock


\bibitem[\protect\citeauthoryear{Inc.}{Inc.}{2021a}]%
        {av2021homepage}
\bibfield{author}{\bibinfo{person}{Alpha~Vantage Inc.}}
  \bibinfo{year}{2021}\natexlab{a}.
\newblock \bibinfo{title}{Alpha Vantage}.
\newblock
\newblock
\urldef\tempurl%
\url{https://www.alphavantage.co/}
\showURL{%
\tempurl}


\bibitem[\protect\citeauthoryear{Inc.}{Inc.}{2021b}]%
        {av2021docs}
\bibfield{author}{\bibinfo{person}{Alpha~Vantage Inc.}}
  \bibinfo{year}{2021}\natexlab{b}.
\newblock \bibinfo{title}{Alpha Vantage API Documentation: Intraday (Extended
  History)}.
\newblock
\newblock
\urldef\tempurl%
\url{https://www.alphavantage.co/documentation/#intraday}
\showURL{%
\tempurl}


\bibitem[\protect\citeauthoryear{Jacob~Leal, Napoletano, Roventini, and
  Fagiolo}{Jacob~Leal et~al\mbox{.}}{2016}]%
        {jacob2016rock}
\bibfield{author}{\bibinfo{person}{Sandrine Jacob~Leal}, \bibinfo{person}{Mauro
  Napoletano}, \bibinfo{person}{Andrea Roventini}, {and}
  \bibinfo{person}{Giorgio Fagiolo}.} \bibinfo{year}{2016}\natexlab{}.
\newblock \showarticletitle{Rock around the clock: An agent-based model of
  low-and high-frequency trading}.
\newblock \bibinfo{journal}{\emph{Journal of Evolutionary Economics}}
  \bibinfo{volume}{26} (\bibinfo{year}{2016}), \bibinfo{pages}{49--76}.
\newblock


\bibitem[\protect\citeauthoryear{Laine}{Laine}{2015}]%
        {laine2015quantitative}
\bibfield{author}{\bibinfo{person}{Tatu Laine}.}
  \bibinfo{year}{2015}\natexlab{}.
\newblock \showarticletitle{Quantitative analysis of financial market
  infrastructures: further perspectives on financial stability}.
\newblock  (\bibinfo{year}{2015}).
\newblock


\bibitem[\protect\citeauthoryear{le~Calvez and Cliff}{le~Calvez and
  Cliff}{2018}]%
        {le2018deep}
\bibfield{author}{\bibinfo{person}{Arthur le Calvez} {and}
  \bibinfo{person}{Dave Cliff}.} \bibinfo{year}{2018}\natexlab{}.
\newblock \showarticletitle{Deep learning can replicate adaptive traders in a
  limit-order-book financial market}. In \bibinfo{booktitle}{\emph{2018 IEEE
  Symposium Series on Computational Intelligence (SSCI)}}.
  \bibinfo{publisher}{IEEE}, \bibinfo{pages}{1876--1883}.
\newblock


\bibitem[\protect\citeauthoryear{LeBaron}{LeBaron}{2001}]%
        {lebaron2001builder}
\bibfield{author}{\bibinfo{person}{Blake LeBaron}.}
  \bibinfo{year}{2001}\natexlab{}.
\newblock \showarticletitle{A builder's guide to agent-based financial
  markets}.
\newblock \bibinfo{journal}{\emph{Quantitative finance}}  \bibinfo{volume}{1}
  (\bibinfo{year}{2001}), \bibinfo{pages}{254--261}.
\newblock


\bibitem[\protect\citeauthoryear{LeBaron}{LeBaron}{2006}]%
        {lebaron2006agent}
\bibfield{author}{\bibinfo{person}{Blake LeBaron}.}
  \bibinfo{year}{2006}\natexlab{}.
\newblock \showarticletitle{Agent-based computational finance}.
\newblock \bibinfo{journal}{\emph{Handbook of computational economics}}
  \bibinfo{volume}{2} (\bibinfo{year}{2006}), \bibinfo{pages}{1187--1233}.
\newblock


\bibitem[\protect\citeauthoryear{LeBaron}{LeBaron}{2011}]%
        {lebaron2011active}
\bibfield{author}{\bibinfo{person}{Blake LeBaron}.}
  \bibinfo{year}{2011}\natexlab{}.
\newblock \showarticletitle{Active and passive learning in agent-based
  financial markets}.
\newblock \bibinfo{journal}{\emph{Eastern Economic Journal}}
  \bibinfo{volume}{37}, \bibinfo{number}{1} (\bibinfo{year}{2011}),
  \bibinfo{pages}{35--43}.
\newblock


\bibitem[\protect\citeauthoryear{Li, Wang, Lin, Sinha, and Wellman}{Li
  et~al\mbox{.}}{2020}]%
        {li2020generating}
\bibfield{author}{\bibinfo{person}{Junyi Li}, \bibinfo{person}{Xintong Wang},
  \bibinfo{person}{Yaoyang Lin}, \bibinfo{person}{Arunesh Sinha}, {and}
  \bibinfo{person}{Michael Wellman}.} \bibinfo{year}{2020}\natexlab{}.
\newblock \showarticletitle{Generating realistic stock market order streams}.
  In \bibinfo{booktitle}{\emph{Proceedings of the AAAI Conference on Artificial
  Intelligence}}, Vol.~\bibinfo{volume}{34}. \bibinfo{publisher}{AAAI},
  \bibinfo{pages}{727--734}.
\newblock


\bibitem[\protect\citeauthoryear{Liang, Liaw, Moritz, Nishihara, Fox, Goldberg,
  Gonzalez, Jordan, and Stoica}{Liang et~al\mbox{.}}{2017}]%
        {liang2017rllib}
\bibfield{author}{\bibinfo{person}{Eric Liang}, \bibinfo{person}{Richard Liaw},
  \bibinfo{person}{Philipp Moritz}, \bibinfo{person}{Robert Nishihara},
  \bibinfo{person}{Roy Fox}, \bibinfo{person}{Ken Goldberg},
  \bibinfo{person}{Joseph~E Gonzalez}, \bibinfo{person}{Michael~I Jordan},
  {and} \bibinfo{person}{Ion Stoica}.} \bibinfo{year}{2017}\natexlab{}.
\newblock \showarticletitle{RLlib: Abstractions for Distributed Reinforcement
  Learning}.
\newblock \bibinfo{journal}{\emph{arXiv preprint arXiv:1712.09381}}
  (\bibinfo{year}{2017}).
\newblock


\bibitem[\protect\citeauthoryear{Lo}{Lo}{2004}]%
        {lo2004adaptive}
\bibfield{author}{\bibinfo{person}{Andrew~W Lo}.}
  \bibinfo{year}{2004}\natexlab{}.
\newblock \showarticletitle{The adaptive markets hypothesis}.
\newblock \bibinfo{journal}{\emph{The Journal of Portfolio Management}}
  \bibinfo{volume}{30}, \bibinfo{number}{5} (\bibinfo{year}{2004}),
  \bibinfo{pages}{15--29}.
\newblock


\bibitem[\protect\citeauthoryear{Madhavan}{Madhavan}{2000}]%
        {madhavan2000market}
\bibfield{author}{\bibinfo{person}{Ananth Madhavan}.}
  \bibinfo{year}{2000}\natexlab{}.
\newblock \showarticletitle{Market microstructure: A survey}.
\newblock \bibinfo{journal}{\emph{Journal of financial markets}}
  \bibinfo{volume}{3}, \bibinfo{number}{3} (\bibinfo{year}{2000}),
  \bibinfo{pages}{205--258}.
\newblock


\bibitem[\protect\citeauthoryear{Malkiel and Fama}{Malkiel and Fama}{1970}]%
        {malkiel1970efficient}
\bibfield{author}{\bibinfo{person}{Burton~G Malkiel} {and}
  \bibinfo{person}{Eugene~F Fama}.} \bibinfo{year}{1970}\natexlab{}.
\newblock \showarticletitle{Efficient capital markets: A review of theory and
  empirical work}.
\newblock \bibinfo{journal}{\emph{The journal of Finance}}
  \bibinfo{volume}{25}, \bibinfo{number}{2} (\bibinfo{year}{1970}),
  \bibinfo{pages}{383--417}.
\newblock


\bibitem[\protect\citeauthoryear{Marwala}{Marwala}{2017}]%
        {marwala2017rational}
\bibfield{author}{\bibinfo{person}{Tshilidzi Marwala}.}
  \bibinfo{year}{2017}\natexlab{}.
\newblock \showarticletitle{Rational Choice and Artificial Intelligence}.
\newblock \bibinfo{journal}{\emph{arXiv preprint arXiv:1703.10098}}
  (\bibinfo{year}{2017}).
\newblock


\bibitem[\protect\citeauthoryear{McGroarty, Booth, Gerding, and
  Chinthalapati}{McGroarty et~al\mbox{.}}{2019}]%
        {mcgroarty2019high}
\bibfield{author}{\bibinfo{person}{Frank McGroarty}, \bibinfo{person}{Ash
  Booth}, \bibinfo{person}{Enrico Gerding}, {and} \bibinfo{person}{VL~Raju
  Chinthalapati}.} \bibinfo{year}{2019}\natexlab{}.
\newblock \showarticletitle{High frequency trading strategies, market fragility
  and price spikes: an agent based model perspective}.
\newblock \bibinfo{journal}{\emph{Annals of Operations Research}}
  \bibinfo{volume}{282}, \bibinfo{number}{1} (\bibinfo{year}{2019}),
  \bibinfo{pages}{217--244}.
\newblock


\bibitem[\protect\citeauthoryear{Michaels and Osipovich}{Michaels and
  Osipovich}{2020}]%
        {michaels2020pilot}
\bibfield{author}{\bibinfo{person}{Dave Michaels} {and}
  \bibinfo{person}{Alexander Osipovich}.} \bibinfo{year}{2020}\natexlab{}.
\newblock \bibinfo{title}{Appeals Court Rules for Stock Exchanges in Fee Fight
  With SEC}.
\newblock
\newblock
\urldef\tempurl%
\url{https://www.wsj.com/articles/appeals-court-rules-for-stock-exchanges-in-fee-fight-with-sec-11592322391}
\showURL{%
\tempurl}


\bibitem[\protect\citeauthoryear{Networks}{Networks}{2021}]%
        {anova2021llfc}
\bibfield{author}{\bibinfo{person}{Anova~Financial Networks}.}
  \bibinfo{year}{2021}\natexlab{}.
\newblock \bibinfo{title}{Low Latency Financial Connectivity}.
\newblock
\newblock
\urldef\tempurl%
\url{https://anovanetworks.com}
\showURL{%
\tempurl}


\bibitem[\protect\citeauthoryear{O'hara}{O'hara}{1997}]%
        {ohara1997market}
\bibfield{author}{\bibinfo{person}{Maureen O'hara}.}
  \bibinfo{year}{1997}\natexlab{}.
\newblock \bibinfo{booktitle}{\emph{Market microstructure theory}}.
\newblock \bibinfo{publisher}{Wiley}.
\newblock


\bibitem[\protect\citeauthoryear{Paddrik, Hayes, Scherer, and Beling}{Paddrik
  et~al\mbox{.}}{2017}]%
        {paddrik2017effects}
\bibfield{author}{\bibinfo{person}{Mark Paddrik}, \bibinfo{person}{Roy Hayes},
  \bibinfo{person}{William Scherer}, {and} \bibinfo{person}{Peter Beling}.}
  \bibinfo{year}{2017}\natexlab{}.
\newblock \showarticletitle{Effects of limit order book information level on
  market stability metrics}.
\newblock \bibinfo{journal}{\emph{Journal of Economic Interaction and
  Coordination}} \bibinfo{volume}{12}, \bibinfo{number}{2}
  (\bibinfo{year}{2017}), \bibinfo{pages}{221--247}.
\newblock


\bibitem[\protect\citeauthoryear{Pagan}{Pagan}{1996}]%
        {pagan1996econometrics}
\bibfield{author}{\bibinfo{person}{Adrian Pagan}.}
  \bibinfo{year}{1996}\natexlab{}.
\newblock \showarticletitle{The econometrics of financial markets}.
\newblock \bibinfo{journal}{\emph{Journal of empirical finance}}
  \bibinfo{volume}{3}, \bibinfo{number}{1} (\bibinfo{year}{1996}),
  \bibinfo{pages}{15--102}.
\newblock


\bibitem[\protect\citeauthoryear{Potters and Bouchaud}{Potters and
  Bouchaud}{2003}]%
        {potters2003more}
\bibfield{author}{\bibinfo{person}{Marc Potters} {and}
  \bibinfo{person}{Jean-Philippe Bouchaud}.} \bibinfo{year}{2003}\natexlab{}.
\newblock \showarticletitle{More statistical properties of order books and
  price impact}.
\newblock \bibinfo{journal}{\emph{Physica A: Statistical Mechanics and its
  Applications}} \bibinfo{volume}{324}, \bibinfo{number}{1-2}
  (\bibinfo{year}{2003}), \bibinfo{pages}{133--140}.
\newblock


\bibitem[\protect\citeauthoryear{Preist and van Tol}{Preist and van
  Tol}{1998}]%
        {preist1998adaptive}
\bibfield{author}{\bibinfo{person}{Chris Preist} {and} \bibinfo{person}{Maarten
  van Tol}.} \bibinfo{year}{1998}\natexlab{}.
\newblock \showarticletitle{Adaptive agents in a persistent shout double
  auction}. In \bibinfo{booktitle}{\emph{Proceedings of the first international
  conference on Information and computation economies}}.
  \bibinfo{publisher}{ACM}, \bibinfo{pages}{11--18}.
\newblock


\bibitem[\protect\citeauthoryear{Rand and Wilensky}{Rand and Wilensky}{2006}]%
        {rand2006verification}
\bibfield{author}{\bibinfo{person}{William Rand} {and} \bibinfo{person}{Uri
  Wilensky}.} \bibinfo{year}{2006}\natexlab{}.
\newblock \showarticletitle{Verification and validation through replication: A
  case study using Axelrod and Hammond’s ethnocentrism model}.
\newblock \bibinfo{journal}{\emph{North American Association for Computational
  Social and Organization Sciences (NAACSOS)}} (\bibinfo{year}{2006}),
  \bibinfo{pages}{1--6}.
\newblock


\bibitem[\protect\citeauthoryear{Rollins and Cliff}{Rollins and Cliff}{2020}]%
        {rollins2020trading}
\bibfield{author}{\bibinfo{person}{Michael Rollins} {and} \bibinfo{person}{Dave
  Cliff}.} \bibinfo{year}{2020}\natexlab{}.
\newblock \showarticletitle{Which Trading Agent is Best? Using a Threaded
  Parallel Simulation of a Financial Market Changes the Pecking-Order}.
\newblock \bibinfo{journal}{\emph{arXiv preprint arXiv:2009.06905}}
  (\bibinfo{year}{2020}).
\newblock


\bibitem[\protect\citeauthoryear{Sabzian, Shafia, Maleki, Hashemi, Baghaei, and
  Gharib}{Sabzian et~al\mbox{.}}{2019}]%
        {sabzian2019theories}
\bibfield{author}{\bibinfo{person}{Hossein Sabzian},
  \bibinfo{person}{Mohammad~Ali Shafia}, \bibinfo{person}{Ali Maleki},
  \bibinfo{person}{Seyeed Mostapha~Seyeed Hashemi}, \bibinfo{person}{Ali
  Baghaei}, {and} \bibinfo{person}{Hossein Gharib}.}
  \bibinfo{year}{2019}\natexlab{}.
\newblock \showarticletitle{Theories and practice of agent based modeling: Some
  practical implications for economic planners}.
\newblock \bibinfo{journal}{\emph{arXiv preprint arXiv:1901.08932}}
  (\bibinfo{year}{2019}).
\newblock


\bibitem[\protect\citeauthoryear{Schmerken}{Schmerken}{2008}]%
        {schmerken2008alpha}
\bibfield{author}{\bibinfo{person}{Ivy Schmerken}.}
  \bibinfo{year}{2008}\natexlab{}.
\newblock \bibinfo{title}{Quants Demand More Efficient Alpha Generation
  Technology Platform}.
\newblock
\newblock
\urldef\tempurl%
\url{https://web.archive.org/web/20110718225953/http://www.wallstreetandtech.com/asset-management/showArticle.jhtml?articleID=208808533}
\showURL{%
\tempurl}
\newblock
\shownote{Accessed 2021-03-16.}


\bibitem[\protect\citeauthoryear{Schvartzman and Wellman}{Schvartzman and
  Wellman}{2009}]%
        {schvartzman2009stronger}
\bibfield{author}{\bibinfo{person}{L~Julian Schvartzman} {and}
  \bibinfo{person}{Michael~P Wellman}.} \bibinfo{year}{2009}\natexlab{}.
\newblock \showarticletitle{Stronger CDA strategies through empirical
  game-theoretic analysis and reinforcement learning}. In
  \bibinfo{booktitle}{\emph{Proceedings of The 8th International Conference on
  Autonomous Agents and Multiagent Systems-Volume 1}}.
  \bibinfo{publisher}{International Foundation for Autonomous Agents and
  Multiagent Systems}, \bibinfo{pages}{249--256}.
\newblock


\bibitem[\protect\citeauthoryear{Securities and Commission}{Securities and
  Commission}{2005}]%
        {sec2005regnms}
\bibfield{author}{\bibinfo{person}{US Securities} {and}
  \bibinfo{person}{Exchanges Commission}.} \bibinfo{year}{2005}\natexlab{}.
\newblock \showarticletitle{Regulation National Market System}.
\newblock  (\bibinfo{year}{2005}).
\newblock
\urldef\tempurl%
\url{https://www.sec.gov/rules/final/34-51808.pdf}
\showURL{%
\tempurl}


\bibitem[\protect\citeauthoryear{Securities and Commission}{Securities and
  Commission}{2010}]%
        {sec2010flashcrash}
\bibfield{author}{\bibinfo{person}{U.S.\ Securities} {and}
  \bibinfo{person}{Exchanges Commission}.} \bibinfo{year}{2010}\natexlab{}.
\newblock \bibinfo{title}{Findings Regarding The Market Events of May 6, 2010}.
\newblock
\newblock
\urldef\tempurl%
\url{https://www.sec.gov/files/marketevents-report.pdf}
\showURL{%
\tempurl}


\bibitem[\protect\citeauthoryear{Securities and Commission}{Securities and
  Commission}{2021a}]%
        {sec2021public}
\bibfield{author}{\bibinfo{person}{U.S.\ Securities} {and}
  \bibinfo{person}{Exchanges Commission}.} \bibinfo{year}{2021}\natexlab{a}.
\newblock \bibinfo{title}{How to Submit Comments}.
\newblock
\newblock
\urldef\tempurl%
\url{https://www.sec.gov/rules/submitcomments.htm}
\showURL{%
\tempurl}


\bibitem[\protect\citeauthoryear{Securities and Commission}{Securities and
  Commission}{2021b}]%
        {sec2021tickpilot}
\bibfield{author}{\bibinfo{person}{U.S.\ Securities} {and}
  \bibinfo{person}{Exchanges Commission}.} \bibinfo{year}{2021}\natexlab{b}.
\newblock \bibinfo{title}{Tick Size Pilot Program}.
\newblock
\newblock
\urldef\tempurl%
\url{https://www.sec.gov/ticksizepilot}
\showURL{%
\tempurl}


\bibitem[\protect\citeauthoryear{Sewell}{Sewell}{2011}]%
        {sewell2011characterization}
\bibfield{author}{\bibinfo{person}{Martin Sewell}.}
  \bibinfo{year}{2011}\natexlab{}.
\newblock \showarticletitle{Characterization of financial time series}.
\newblock \bibinfo{journal}{\emph{Rn}} \bibinfo{volume}{11},
  \bibinfo{number}{01} (\bibinfo{year}{2011}), \bibinfo{pages}{01}.
\newblock


\bibitem[\protect\citeauthoryear{Sirignano and Cont}{Sirignano and
  Cont}{2019}]%
        {sirignano2019universal}
\bibfield{author}{\bibinfo{person}{Justin Sirignano} {and}
  \bibinfo{person}{Rama Cont}.} \bibinfo{year}{2019}\natexlab{}.
\newblock \showarticletitle{Universal features of price formation in financial
  markets: perspectives from deep learning}.
\newblock \bibinfo{journal}{\emph{Quantitative Finance}} \bibinfo{volume}{19},
  \bibinfo{number}{9} (\bibinfo{year}{2019}), \bibinfo{pages}{1449--1459}.
\newblock


\bibitem[\protect\citeauthoryear{Smith, Farmer, Gillemot, and
  Krishnamurthy}{Smith et~al\mbox{.}}{2003}]%
        {smith2003statistical}
\bibfield{author}{\bibinfo{person}{Eric Smith}, \bibinfo{person}{J~Doyne
  Farmer}, \bibinfo{person}{L{\'a}szl{\'o} Gillemot}, {and}
  \bibinfo{person}{Supriya Krishnamurthy}.} \bibinfo{year}{2003}\natexlab{}.
\newblock \showarticletitle{Statistical theory of the continuous double
  auction}.
\newblock \bibinfo{journal}{\emph{Quantitative finance}}  \bibinfo{volume}{3}
  (\bibinfo{year}{2003}), \bibinfo{pages}{481--514}.
\newblock


\bibitem[\protect\citeauthoryear{Soram{\"a}ki, Bech, Arnold, Glass, and
  Beyeler}{Soram{\"a}ki et~al\mbox{.}}{2006}]%
        {soramaki2006topology}
\bibfield{author}{\bibinfo{person}{Kimmo Soram{\"a}ki},
  \bibinfo{person}{Morten~L Bech}, \bibinfo{person}{Jeffrey Arnold},
  \bibinfo{person}{Robert~J Glass}, {and} \bibinfo{person}{Walter~E Beyeler}.}
  \bibinfo{year}{2006}\natexlab{}.
\newblock \showarticletitle{The Topology of Interbank Payment Flows}.
\newblock \bibinfo{journal}{\emph{Federal Reserve Bank of New York: Staff
  Reports}} (\bibinfo{year}{2006}).
\newblock


\bibitem[\protect\citeauthoryear{Sorensen, Ozzello, Rogan, Baker, Parks, and
  Hu}{Sorensen et~al\mbox{.}}{2020}]%
        {sorensen2020meta}
\bibfield{author}{\bibinfo{person}{Erik Sorensen}, \bibinfo{person}{Ryan
  Ozzello}, \bibinfo{person}{Rachael Rogan}, \bibinfo{person}{Ethan Baker},
  \bibinfo{person}{Nate Parks}, {and} \bibinfo{person}{Wei Hu}.}
  \bibinfo{year}{2020}\natexlab{}.
\newblock \showarticletitle{Meta-Learning of Evolutionary Strategy for Stock
  Trading}.
\newblock \bibinfo{journal}{\emph{Journal of Data Analysis and Information
  Processing}} \bibinfo{volume}{8}, \bibinfo{number}{2} (\bibinfo{year}{2020}),
  \bibinfo{pages}{86--98}.
\newblock


\bibitem[\protect\citeauthoryear{Stotter, Cartlidge, and Cliff}{Stotter
  et~al\mbox{.}}{2014}]%
        {stotter2014behavioural}
\bibfield{author}{\bibinfo{person}{Steve Stotter}, \bibinfo{person}{John
  Cartlidge}, {and} \bibinfo{person}{Dave Cliff}.}
  \bibinfo{year}{2014}\natexlab{}.
\newblock \showarticletitle{Behavioural investigations of financial trading
  agents using Exchange Portal (ExPo)}.
\newblock In \bibinfo{booktitle}{\emph{Transactions on Computational Collective
  Intelligence XVII}}. \bibinfo{publisher}{Springer}, \bibinfo{pages}{22--45}.
\newblock


\bibitem[\protect\citeauthoryear{Subramanian, Ramamoorthy, Stone, and
  Kuipers}{Subramanian et~al\mbox{.}}{2006}]%
        {subramanian2006designing}
\bibfield{author}{\bibinfo{person}{Harish Subramanian},
  \bibinfo{person}{Subramanian Ramamoorthy}, \bibinfo{person}{Peter Stone},
  {and} \bibinfo{person}{Benjamin~J Kuipers}.} \bibinfo{year}{2006}\natexlab{}.
\newblock \showarticletitle{Designing safe, profitable automated stock trading
  agents using evolutionary algorithms}. In
  \bibinfo{booktitle}{\emph{Proceedings of the 8th annual conference on Genetic
  and evolutionary computation}}. \bibinfo{publisher}{ACM},
  \bibinfo{pages}{1777--1784}.
\newblock


\bibitem[\protect\citeauthoryear{Talla~Kuate}{Talla~Kuate}{2016}]%
        {talla2016hierarchical}
\bibfield{author}{\bibinfo{person}{Rodrigue Talla~Kuate}.}
  \bibinfo{year}{2016}\natexlab{}.
\newblock \emph{\bibinfo{title}{Hierarchical reinforcement learning for trading
  agents}}.
\newblock \bibinfo{thesistype}{Ph.D. Dissertation}. \bibinfo{school}{Aston
  University}.
\newblock


\bibitem[\protect\citeauthoryear{Tesauro and Bredin}{Tesauro and
  Bredin}{2002}]%
        {tesauro2002strategic}
\bibfield{author}{\bibinfo{person}{Gerald Tesauro} {and}
  \bibinfo{person}{Jonathan~L Bredin}.} \bibinfo{year}{2002}\natexlab{}.
\newblock \showarticletitle{Strategic sequential bidding in auctions using
  dynamic programming}. In \bibinfo{booktitle}{\emph{Proceedings of the first
  international joint conference on Autonomous agents and multiagent systems:
  part 2}}. \bibinfo{publisher}{International Foundation for Autonomous Agents
  and Multiagent Systems}, \bibinfo{pages}{591--598}.
\newblock


\bibitem[\protect\citeauthoryear{Tesauro and Das}{Tesauro and Das}{2001}]%
        {tesauro2001high}
\bibfield{author}{\bibinfo{person}{Gerald Tesauro} {and}
  \bibinfo{person}{Rajarshi Das}.} \bibinfo{year}{2001}\natexlab{}.
\newblock \showarticletitle{High-performance bidding agents for the continuous
  double auction}. In \bibinfo{booktitle}{\emph{Proceedings of the 3rd ACM
  Conference on Electronic Commerce}}. \bibinfo{publisher}{ACM},
  \bibinfo{pages}{206--209}.
\newblock


\bibitem[\protect\citeauthoryear{Tivnan, Dewhurst, Van~Oort, Ring~IV, Gray,
  Tivnan, Koehler, McMahon, Slater, Veneman, et~al\mbox{.}}{Tivnan
  et~al\mbox{.}}{2020}]%
        {tivnan2020fragmentation}
\bibfield{author}{\bibinfo{person}{Brian~F Tivnan},
  \bibinfo{person}{David~Rushing Dewhurst}, \bibinfo{person}{Colin~M Van~Oort},
  \bibinfo{person}{John~H Ring~IV}, \bibinfo{person}{Tyler~J Gray},
  \bibinfo{person}{Brendan~F Tivnan}, \bibinfo{person}{Matthew~TK Koehler},
  \bibinfo{person}{Matthew~T McMahon}, \bibinfo{person}{David~M Slater},
  \bibinfo{person}{Jason~G Veneman}, {et~al\mbox{.}}}
  \bibinfo{year}{2020}\natexlab{}.
\newblock \showarticletitle{Fragmentation and inefficiencies in US equity
  markets: Evidence from the Dow 30}.
\newblock \bibinfo{journal}{\emph{PloS one}} \bibinfo{volume}{15},
  \bibinfo{number}{1} (\bibinfo{year}{2020}), \bibinfo{pages}{e0226968}.
\newblock


\bibitem[\protect\citeauthoryear{Tivnan, Koehler, Slater, Veneman, and
  Tivnan}{Tivnan et~al\mbox{.}}{2017}]%
        {tivnan2017sip}
\bibfield{author}{\bibinfo{person}{Brian~F Tivnan}, \bibinfo{person}{Matthew~TK
  Koehler}, \bibinfo{person}{David Slater}, \bibinfo{person}{Jason Veneman},
  {and} \bibinfo{person}{Brendan~F Tivnan}.} \bibinfo{year}{2017}\natexlab{}.
\newblock \showarticletitle{Towards a model of the US stock market: How
  important is the securities information processor?}. In
  \bibinfo{booktitle}{\emph{2017 Winter Simulation Conference (WSC)}}. IEEE,
  \bibinfo{pages}{1181--1192}.
\newblock


\bibitem[\protect\citeauthoryear{Van~Oort}{Van~Oort}{2018}]%
        {vanoort2018market}
\bibfield{author}{\bibinfo{person}{Colin~M. Van~Oort}.}
  \bibinfo{year}{2018}\natexlab{}.
\newblock \showarticletitle{Market Efficiency in US Stock Markets: A Study of
  the Dow 30 and the S\&P 30}.
\newblock \bibinfo{journal}{\emph{Graduate College Dissertations and Theses}}
  (\bibinfo{year}{2018}).
\newblock


\bibitem[\protect\citeauthoryear{Van~Oort}{Van~Oort}{2021}]%
        {vanoort2021abmms}
\bibfield{author}{\bibinfo{person}{Colin~M. Van~Oort}.}
  \bibinfo{year}{2021}\natexlab{}.
\newblock \bibinfo{title}{Agent Based Market Microstructure Simulation}.
\newblock
\newblock
\urldef\tempurl%
\url{https://gitlab.com/computational-finance-lab/abmms}
\showURL{%
\tempurl}
\newblock
\shownote{Accessed 2021/04/20.}


\bibitem[\protect\citeauthoryear{Vives}{Vives}{1993}]%
        {vives1993fast}
\bibfield{author}{\bibinfo{person}{Xavier Vives}.}
  \bibinfo{year}{1993}\natexlab{}.
\newblock \showarticletitle{How fast do rational agents learn?}
\newblock \bibinfo{journal}{\emph{The Review of Economic Studies}}
  \bibinfo{volume}{60}, \bibinfo{number}{2} (\bibinfo{year}{1993}),
  \bibinfo{pages}{329--347}.
\newblock


\bibitem[\protect\citeauthoryear{Vytelingum, Cliff, and Jennings}{Vytelingum
  et~al\mbox{.}}{2008}]%
        {vytelingum2008strategic}
\bibfield{author}{\bibinfo{person}{Perukrishnen Vytelingum},
  \bibinfo{person}{Dave Cliff}, {and} \bibinfo{person}{Nicholas~R Jennings}.}
  \bibinfo{year}{2008}\natexlab{}.
\newblock \showarticletitle{Strategic bidding in continuous double auctions}.
\newblock \bibinfo{journal}{\emph{Artificial Intelligence}}
  \bibinfo{volume}{172}, \bibinfo{number}{14} (\bibinfo{year}{2008}),
  \bibinfo{pages}{1700--1729}.
\newblock


\bibitem[\protect\citeauthoryear{Wah and Wellman}{Wah and Wellman}{2016}]%
        {wah2016latency}
\bibfield{author}{\bibinfo{person}{Elaine Wah} {and} \bibinfo{person}{Michael~P
  Wellman}.} \bibinfo{year}{2016}\natexlab{}.
\newblock \showarticletitle{Latency arbitrage in fragmented markets: A
  strategic agent-based analysis}.
\newblock \bibinfo{journal}{\emph{Algorithmic Finance}} \bibinfo{volume}{5},
  \bibinfo{number}{3-4} (\bibinfo{year}{2016}), \bibinfo{pages}{69--93}.
\newblock


\bibitem[\protect\citeauthoryear{Wah, Wright, and Wellman}{Wah
  et~al\mbox{.}}{2017}]%
        {wah2017welfare}
\bibfield{author}{\bibinfo{person}{Elaine Wah}, \bibinfo{person}{Mason Wright},
  {and} \bibinfo{person}{Michael~P Wellman}.} \bibinfo{year}{2017}\natexlab{}.
\newblock \showarticletitle{Welfare effects of market making in continuous
  double auctions}.
\newblock \bibinfo{journal}{\emph{Journal of Artificial Intelligence Research}}
   \bibinfo{volume}{59} (\bibinfo{year}{2017}), \bibinfo{pages}{613--650}.
\newblock


\bibitem[\protect\citeauthoryear{Wang, Kurth-Nelson, Tirumala, Soyer, Leibo,
  Munos, Blundell, Kumaran, and Botvinick}{Wang et~al\mbox{.}}{2016}]%
        {wang2016learning}
\bibfield{author}{\bibinfo{person}{Jane~X Wang}, \bibinfo{person}{Zeb
  Kurth-Nelson}, \bibinfo{person}{Dhruva Tirumala}, \bibinfo{person}{Hubert
  Soyer}, \bibinfo{person}{Joel~Z Leibo}, \bibinfo{person}{Remi Munos},
  \bibinfo{person}{Charles Blundell}, \bibinfo{person}{Dharshan Kumaran}, {and}
  \bibinfo{person}{Matt Botvinick}.} \bibinfo{year}{2016}\natexlab{}.
\newblock \showarticletitle{Learning to reinforcement learn}.
\newblock \bibinfo{journal}{\emph{arXiv preprint arXiv:1611.05763}}
  (\bibinfo{year}{2016}).
\newblock


\bibitem[\protect\citeauthoryear{Wang and Wellman}{Wang and Wellman}{2017}]%
        {wang2017spoofing}
\bibfield{author}{\bibinfo{person}{Xintong Wang} {and}
  \bibinfo{person}{Michael~Paul Wellman}.} \bibinfo{year}{2017}\natexlab{}.
\newblock \showarticletitle{Spoofing the limit order book: An agent-based
  model}. In \bibinfo{booktitle}{\emph{Workshops at the Thirty-First AAAI
  Conference on Artificial Intelligence}}.
\newblock


\bibitem[\protect\citeauthoryear{Wray, Meades, and Cliff}{Wray
  et~al\mbox{.}}{2020}]%
        {wray2020automated}
\bibfield{author}{\bibinfo{person}{Aaron Wray}, \bibinfo{person}{Matthew
  Meades}, {and} \bibinfo{person}{Dave Cliff}.}
  \bibinfo{year}{2020}\natexlab{}.
\newblock \showarticletitle{Automated Creation of a High-Performing Algorithmic
  Trader via Deep Learning on Level-2 Limit Order Book Data}. In
  \bibinfo{booktitle}{\emph{2020 IEEE Symposium Series on Computational
  Intelligence (SSCI)}}. \bibinfo{publisher}{IEEE},
  \bibinfo{pages}{1067--1074}.
\newblock


\bibitem[\protect\citeauthoryear{Xiang, Kennedy, Madey, and Cabaniss}{Xiang
  et~al\mbox{.}}{2005}]%
        {xiang2005verification}
\bibfield{author}{\bibinfo{person}{Xiaorong Xiang}, \bibinfo{person}{Ryan
  Kennedy}, \bibinfo{person}{Gregory Madey}, {and} \bibinfo{person}{Steve
  Cabaniss}.} \bibinfo{year}{2005}\natexlab{}.
\newblock \showarticletitle{Verification and validation of agent-based
  scientific simulation models}. In \bibinfo{booktitle}{\emph{Agent-directed
  simulation conference}}, Vol.~\bibinfo{volume}{47}. \bibinfo{publisher}{The
  European Modeling and Simulation Symposium}, \bibinfo{pages}{55}.
\newblock


\end{thebibliography}

\clearpage
\appendix

\setcounter{figure}{0}
\setcounter{table}{0}
\renewcommand{\thefigure}{S\arabic{figure}}
\renewcommand{\thetable}{S\arabic{table}}

\section{ODD Protocol for ABMMS}\label{sec:odd}
Below we describe ABMMS following the Overview, Design concepts, Details (ODD) protocol~\cite{grimm2006standard, grimm2010odd, grimm2020odd}.

\subsection{Purpose}\label{subsec:purpose}
The immediate goal of ABMMS is to evaluate the impact of market microstructure details and adaptive agents on the quality of data generated by an agent based financial market (ABFM).
Phrased more explicitly, ``Does the combination of detailed market mechanisms and adaptive learning agents create synergistic effects that improve the level of realism of data generated by an ABFM?''
Beyond that, ABMMS is intended as a tool to evaluate the impacts of policy and design decisions in the US National Market System for equities (NMS).
ABMMS primarily targets the NMS, but is designed to simulate arbitrary market configurations to allow for the investigation of a broad range of counterfactual scenarios.

\subsection{Patterns}\label{subsec:patterns}
We evaluate ABMMS by its ability to reproduce the following patterns:

\subsubsection{Stylized Facts of Asset Prices}\label{subsubsec:stylized_facts}
ABMMS should produce asset price time series that satisfy stylized facts proposed by ~\citet{cont2001empirical} and others.
Replicating all of the 11 stylized facts proposed by ~\citet{cont2001empirical} is difficult, especially considering the volume of data that is required to evaluate facts \#1, \#4, and \#11, so we aim to replicate at least 4.
Additionally, we found that facts \#3 and \#10 were difficult to detect when calibrating our stylized fact tests on real data.
Therefore, replicating 4 stylized facts should be considered acceptable and replicating 6 stylized facts should be considered desirable.

\subsubsection{Profits of Learning Agents}\label{subsubsec:learner_profit}
Simple agents may have positive or negative profits depending largely on the market conditions that they experience and random chance.
However, learning agents should have positive expected profits, otherwise economic rationality would demand that they cease participation.
This does not require that an agent have positive profit in any particular time window, or even over the entirety of a particular simulation run, only that an agent nets positive profit in the long run.

\subsubsection{Daily Trading Activity}\label{subsubsec:daily_trading_activity}
Financial markets tend to feature a smile shaped activity density curve at the trading day time scale.
The start of the trading day features a burst of trading activity, which decays as the day goes on, as well as a ramp up of activity as the day reaches its close.
There is no mechanism to generate such an activity curve when an ABFM is populated entirely with ZI or ZIP agents, beyond engineering such activity patterns into their trading behavior.
However, when learning agents are introduced one possible mechanism for generating this activity distribution comes with them, and that is the opportunity cost associated with the market closure between trading days.
To test for this pattern we can construct activity histograms for each trading day, with bins covering 10 second intervals, then test if the bins in the first and last 5 minutes of the trading day feature significantly more activity than other bins.

% \subsubsection{Order Book Metrics}\label{subsubsec:order_book_metrics}
% \todo{CVO: LOB based market metrics~\cite{paddrik2017effects} can be used to quantify the impacts of learning agents at a finer level. How can we turn this into a testable pattern?}

\subsubsection{Dislocations}\label{subsubsec:dislocations}
\citet{tivnan2020fragmentation} describe the occurrence of quote dislocations in the NMS\@.
Since ABMMS is intended to model the NMS, we expect to observe similar quote dislocations in the data generated by it.
The quote dislocations observed in ABMMS should have similar distributions of attributes to what was observed in the NMS\@.
On average, stocks in the NMS can exhibit daily dislocation counts that fall anywhere between roughly 3000 and 16000, for thinly traded members of the Russell 3000 and members of the Dow 30 respectively~\cite{dewhurst2019scaling}.
When accounting for time of day, the occurrence distribution should have a smile-like shape, where more dislocations occur near the open and close of a trading day.
Additionally, the duration distribution should be heavy tailed, with a tail reaching towards longer duration, and a mean between $10^{-4}$ and $10^{-2}$ $\mu$s.
The distribution of dislocation magnitudes should be heavy tailed, possibly a power law, with a greater frequency of small magnitude dislocations and an exceptionally long tail~\cite{vanoort2018market}.

\subsection{Entities}\label{subsec:entities}
ABMMS features the following:
\begin{itemize}
    \item Simulation Driver: Orchestrates the execution of the simulation and manages global variables.
   
    \item Communication Network (CN): Mediates interactions between agents.
    
    \item Agent: An actor in the simulation.
    Agent classes can have heterogeneous roles and incentives.
    All agents share a set of common state variables that cover general information.
    \begin{itemize}
        \item Exchange: Manage auctions that facilitate stock trading.
        \item Securities Information Processor (SIP): Provides a signal to synchronize prices across a fragmented marketplace.
        \begin{itemize}
            \item Limit Up-Limit Down Queue: Tracks historical trades in a time window to aid in calculating LULD bands.
        \end{itemize}
        \item Trader: Buys and sells financial instruments.
        \begin{itemize}
            \item Zero Intelligence (ZI):
            Based on the agents developed by ~\citet{gode1993allocative}, all trading decisions are selected randomly.
            One deviation is that we do not implement separate buyer and seller agents. Instead, each time a ZI agent is able to trade it randomly selects whether to act as a buyer or seller.
            \item Minimum Intelligence (MI):
            Similar to ZI agents, but the width of the random price distribution is based on the spread of the NBBO, and orders are always sent to the exchange that holds the appropriate side of the NBBO.
            \item ZI Plus (ZIP):
            Based on the agents developed by ~\citet{cliff1997zero}.
            Takes trading actions that are nearly as random as ZI agents, except that prices are determined by a belief function that is updated based on orders that result in trades.
            \item Arbitrage:
            Attempts to profit by uncrossing distributed markets that are crossed.
            \item Reinforcement Learning (RL):
            Learns a trading strategy via meta-reinforcement learning, leading to a more adaptive and free form strategy.
        \end{itemize}
        \item Observer: An aggregator that constructs consolidated data products.
    \end{itemize}
    
    \item Message: Information sent from one agent to another.
    All messages feature the same header information, while the body content varies based on the message type.
    \begin{itemize}
        \item Add: A bid (buy interest) or offer (sell interest) has been added to an order book.
        \item Modify (Mod): Shares have been removed from an order book without execution.
        \item Trade: Shares have been removed from an order book due to execution.
        \item Quote: The best bid or best offer at an exchange have updated.
        \item National Best Bid and Offer (NBBO): The best bid or best offer across all exchanges in a market system have updated.
        \item Limit Up-Limit Down (LULD) Bands: Range of valid trading prices for an asset.
        \item Request: Traders may submit an add or mod request to an exchange. Requests may be rejected if they are malformed.
        \item Receipt: Exchanges indicate the status of a request via a receipt that is sent exclusively to the sender of the request.
        \item Trigger: Schedules the occurrence of a discrete event, such as a trade or an auction. Usually sent from an agent to itself, though this is not explicitly enforced.
        \item SIP Message: Trade and Quote messages that pass through a SIP.
    \end{itemize}
\end{itemize}

CNs represent the communication infrastructure through which all other agents interact.
The core of an CN is the topology of the communication infrastructure, which is represented as an undirected graph.
The nodes of the topology represent physical locations, and edges represent communication pathways between locations.
Edges are weighted to represent a deterministic propagation delay associated with sending a message across that edge.
An exponential random variable is added to the deterministic propagation delay, simulating other aspects of electronic communication systems, such as queuing delays or packet loss.
All messages sent via the CN are subjected to a minimum delay, which primarily impacts messages sent between agents located at the same node.
The state variables for CNs are summarized in Table~\ref{tab:CN_sv}.

Exchanges facilitate the trade of assets by matching buyers and sellers via an auction mechanism.
The auction mechanism is implemented by the combination of an order book, which accumulates market state, and a matching engine, which matches incoming orders against those resting in the order book.
Trading in multiple assets can be supported through the use of multiple independent order books.
Exchanges may use transaction fees, also called market access fees or maker-taker fees, to monetize their activity.
The state variables for exchanges are summarized in Table~\ref{tab:exchange_sv}.

SIPs act as a synchronization mechanism by aggregating information across a fragmented market system and disseminating indicators.
A SIP constructs several signals, including the national best bid and offer (NBBO), limit up-limit down (LULD) band indicators, as well as a trade and quote (TAQ) for each asset it is responsible for.
The state variables for SIPs are summarized in Table~\ref{tab:sip_sv}.
Each SIP tracks historical trades over a small time window in order to implement the Limit Up-Limit Down (LULD) mechanism.
These trades are stored in a LULD Queue, which aids in the calculation of LULD bands.
The state variables for LULD Queues are summarized in Table~\ref{tab:luld_queue_sv}.

Traders buy, sell, and hold financial instruments by interacting with other traders via an exchange.
Each trader tracks the state of its holdings, the amount of each traded asset, plus cash, that it possesses.
Additionally, each trader implements a strategy for placing bids and offers.
The state variables for Traders are summarized in Tables~\ref{tab:trader_sv}--\ref{tab:arb_sv}.

\begin{figure*}
    \centering
    \includegraphics[width=0.94\textwidth]{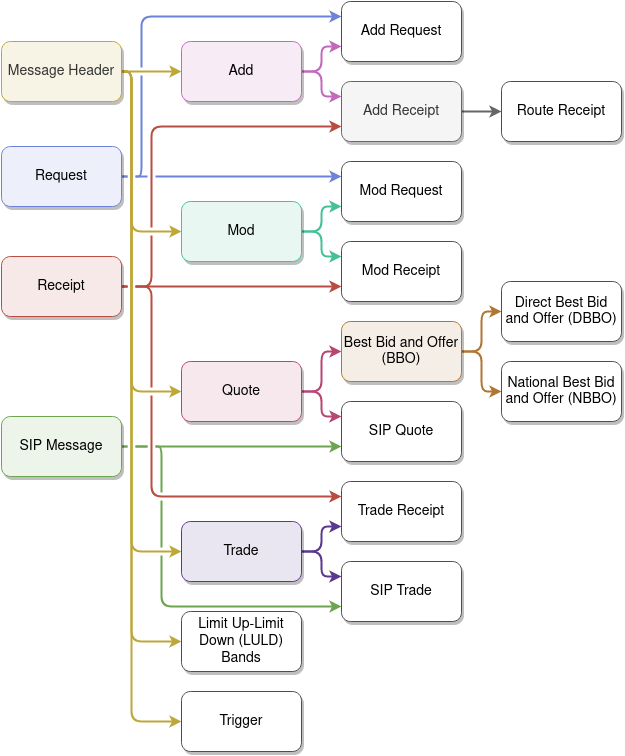}
    \caption{
        A graphical summary of the relationships between the message types implemented in ABMMS\@.
        Message types that are higher up in the tree share their state variables with message types that are lower in the tree, if they are connected.
    }
    \label{fig:message_tree}
\end{figure*}

ABMMS implements a variety of message types, whose relationships are summarized in Figure~\ref{fig:message_tree}.
The state variables for each message type are summarized in Tables~\ref{tab:message_header_sv}--\ref{tab:trigger_sv}.

\subsection{State Variables}\label{subsec:state_vars}
\begin{table*}[htp]
    \centering
    \caption{State variables for the Simulation Driver entity.}
    \begin{tabularx}{\textwidth}{llX}
        Variable Name & Variable Type and Units & Meaning \\
        \hline
        Global Clock & Timestamp, dynamic; $\mu$s & Time keeper for the simulation. \\
        Simulation Start & Timestamp, static; $\mu$s & The Global Clock is set to this at the start of the simulation. \\
        Simulation End Time & Timestamp, static; $\mu$s & The simulation is terminated if the Global Clock reaches or passes this. \\
        Communication Network & CN, static & Communication infrastructure that mediates agent interactions. See Table~\ref{tab:CN_sv} for more details. \\
        Agents & List[Agent], static & Agents that populate this simulation. See Table~\ref{tab:agent_sv} and related Tables for more details. \\
        Trading Symbols & List[String], static & Identifiers for the stocks that will be traded in this simulation. \\
    \end{tabularx}
    \label{tab:driver_sv}
\end{table*}

\begin{table*}[htp]
    \centering
    \caption{State variables for the Communication Network (CN) entity.}
    \begin{tabularx}{\textwidth}{llX}
        Variable Name & Variable Type and Units & Meaning \\
        \hline
        Topology & Undirected Graph, static & Nodes represent physical locations that other agents might inhabit. Edges represent communication channels between locations. Weights on edges indicate the magnitude of deterministic propagation delays associated with communication along each edge. \\
        Minimum Delay & Integer, static; $\mu$s & The minimum delay imposed on all communications. Primarily impacts messages passed between agents located at the same node in the CN\@. \\
        Mean Delay Noise & Float, static; $\mu$s & Scale parameter for an exponential random variable that is used to create stochastic communication delays. \\
        Message Queue & Sorted Queue, dynamic & Contains the messages that have been sent into the CN, but not yet arrived. Always sorted such that the first element of the queue is the next message that will arrive at its destination. \\
    \end{tabularx}
    \label{tab:CN_sv}
\end{table*}

\begin{table*}[htp]
    \centering
    \caption{State variables for the Agent entity.}
    \begin{tabularx}{\textwidth}{llX}
        Variable Name & Variable Type and Units & Meaning \\
        \hline
        Identifier & String, static & A unique identifier, or name, that is used to refer to this agent. \\
        Location & Categorical, static & A node in the CN where this agent is located. \\
        Clock & Timestamp, dynamic; $\mu$s & A local clock. A copy of the global simulation clock by default.  \\
        Trading Symbols & List[String], static  & Identifiers of stocks that this agent may interact with. \\
        Subscribers & List[Agent], dynamic & Agents subscribed to the broadcast feed of this agent. \\
    \end{tabularx}
    \label{tab:agent_sv}
\end{table*}

\begin{table*}[htp]
    \centering
    \caption{State variables for the Exchange entity.}
    \begin{tabularx}{\textwidth}{llX}
        Variable Name & Variable Type and Units & Meaning \\
        \hline
        Agent State Variables & N/A & See Table~\ref{tab:agent_sv} for more details. \\
        Order Books & Dict[String, Order Book], static & Mapping from Trading Symbols to their associated Order books (Table~\ref{tab:order_book_sv}). \\
        Matching Engine & Matching Engine, static & Strategy for matching incoming orders with resting orders. \\
    \end{tabularx}
    \label{tab:exchange_sv}
\end{table*}

\begin{table*}[htp]
    \centering
    \caption{State variables for the Order Book entity.}
    \begin{tabularx}{\textwidth}{llX}
        Variable Name & Variable Type and Units & Meaning \\
        \hline
        Trading Symbol & String, static & Orders are managed for this trading symbol. \\
        Bid Priority & Ordering Function, static & Defines an ordering for the execution priority of bids. \\
        Bids & List[Add Request], dynamic & List of Bid Requests that have been accepted, but not executed. Sorted according to Bid Priority. \\
        Offer Priority & Ordering Function, static & Defines an ordering for the execution priority of offers. \\
        Offers & List[Add Request], dynamic & List of Offer Requests that have been accepted, but not executed. Sorted according to Offer Priority. \\
    \end{tabularx}
    \label{tab:order_book_sv}
\end{table*}

\begin{table*}[htp]
    \centering
    \caption{State variables for the Securities Information Processor (SIP) entity.}
    \begin{tabularx}{\textwidth}{llX}
        Variable Name & Variable Type and Units & Meaning \\
        \hline
        Agent State Variables & N/A & See Table~\ref{tab:agent_sv} for more details. \\
        LULD Queues & Dict[String, LULD Queue], static & Map from trading symbols to LULD Queues. One LULD Queue for each trading symbol that this SIP is responsible for. See Table~\ref{tab:luld_queue_sv} for more details. \\
        Round Lot Size & Integer, static; shares of stock & How many shares must be associated with a quote for it to be considered a round lot, and thus eligible for inclusion in the NBBO. \\
    \end{tabularx}
    \label{tab:sip_sv}
\end{table*}

\begin{table*}[htp]
    \centering
    \caption{State variables for the Limit Up-Limit Down (LULD) Queue entity.}
    \begin{tabularx}{\textwidth}{llX}
        Variable Name & Variable Type and Units & Meaning \\
        \hline
        Trading Symbol & String, static & LULD bands are managed for this trading symbol. \\
        LULD Reference Price & Integer, static & Initial reference price for the LULD bands. \\
        LULD Window & Timedelta, static; $\mu$s & Length of the time window used to select recent trades. \\
        LULD Percentage & Float, static; percent & Half the width of the LULD bands as a fraction of the reference price. \\
    \end{tabularx}
    \label{tab:luld_queue_sv}
\end{table*}

\begin{table*}[htp]
    \centering
    \caption{State variables for the Trader entity.}
    \begin{tabularx}{\textwidth}{lXX}
        Variable Name & Variable Type and Units & Meaning \\
        \hline
        Agent State Variables & N/A & See Table~\ref{tab:agent_sv} for more details. \\
        Holdings & Dict[String, Integer or Float], dynamic; Shares of stock or \$0.0001 & Mapping from asset identifiers to possessed asset quantities. \\
        Pending Orders & Dict[Integer, Message], dynamic & Mapping from order identifiers to orders that have been submitted to an exchange and have an unknown status. \\
        Active Orders & Dict[Integer, Message], dynamic & Mapping from order identifiers to orders that have been placed into an order book on an exchange. \\
        NBBOs & Dict[String, NBBO], dynamic & Mapping from trading symbols to the current NBBO for that trading symbol. \\
    \end{tabularx}
    \label{tab:trader_sv}
\end{table*}

\begin{table*}[htp]
    \centering
    \caption{
        State variables for the Zero Intelligence (ZI) Trader and Minimum Intelligence (MI) Trader entities.
    }
    \begin{tabularx}{\textwidth}{llX}
        Variable Name & Variable Type and Units & Meaning \\
        \hline
        Trader State Variables & N/A & See Table~\ref{tab:trader_sv} for more details. \\
        Maximum Limit Prices & Dict[String, Integer], dynamic; \$0.01 & Maximum prices for submitted limit orders, one for each traded stock. \\
        Minimum Limit Prices & Dict[String, Integer], dynamic; \$0.01 & Minimum prices for submitted limit orders, one for each traded stock. \\
    \end{tabularx}
    \label{tab:zi_sv}
\end{table*}

\begin{table*}[htp]
    \centering
    \caption{
        State variables for the Zero Intelligence Plus (ZIP) Trader entity.
    }
    \begin{tabularx}{\textwidth}{llX}
        Variable Name & Variable Type and Units & Meaning \\
        \hline
        Trader State Variables & N/A & See Table~\ref{tab:trader_sv} for more details. \\
        Profit Margins & Dict[String, List[Float]], dynamic & Mapping from trading symbols to pairs of profit margins, one for bids and one for offers. \\
        Limit Prices & Dict[String, Integer], dynamic & Mapping from trading symbols to the worst price that the agent is willing to transact at. \\
        Target Prices & Dict[String, List[Integer]], dynamic & Mapping from trading symbols to target prices used to update the profit margins. \\
        Momentum Values & Dict[String, Float], dynamic & Mapping from trading symbols to current momentum values. \\
    \end{tabularx}
    \label{tab:zip_sv}
\end{table*}

\begin{table*}[htp]
    \centering
    \caption{State variables for the Arbitrage Trader and Reinforcement Learning Trader entities.}
    \begin{tabularx}{\textwidth}{llX}
        Variable Name & Variable Type and Units & Meaning \\
        \hline
        Trader State Variables & N/A & See Table~\ref{tab:trader_sv} for more details. \\
        DBBOs & Dict[String, DBBO], dynamic & Mapping from Trading Symbols to their current Direct Best Bid and Offer (DBBO). \\
    \end{tabularx}
    \label{tab:arb_sv}
\end{table*}

\begin{table*}[htp]
    \centering
    \caption{State variables for the Message Header entity.}
    \begin{tabularx}{\textwidth}{llX}
        Variable Name & Variable Type and Units & Meaning \\
        \hline
        Message ID & Integer, static & Identifier associated with this message. \\
        Related ID & Integer, static & Optional identifier of a related message. \\
        Sender ID & String, static & Identifier of the agent that sent the message. \\
        Recipient ID & String, static & Identifier of the intended recipient. \\
        Send Time & Timestamp, static; $\mu$s & When the message was sent. \\
        Receive Time & Timestamp, static; $\mu$s & When the message will be received. \\
        Trading Symbol & String, static & Indicates what Trading Symbol this message is associated with. \\
        Random & Float, static & Value drawn from a $\mathcal{U}[0, 1)$ distribution. \\
    \end{tabularx}
    \label{tab:message_header_sv}
\end{table*}

\begin{table*}[htp]
    \centering
    \caption{State variables for the Add message entity.}
    \begin{tabularx}{\textwidth}{llX}
        Variable Name & Variable Type and Units & Meaning \\
        \hline
        Message Header & N/A & See Table~\ref{tab:message_header_sv} for details. \\
        Sequence Number & Integer, static & Identifier applied by an exchange to indicate the processing order of requests. \\
        Order Type & Categorical, static & Limit, market, or midpoint order. \\
        Side & Categorical, static & Bid or offer. \\
        Shares & Integer, static & Quantity of shares to be bought or sold. \\
        Limit Price & Integer, static; \$0.01 & Highest acceptable bid price, or lowest acceptable offer price. \\
        All or Nothing & Boolean, static & Indicates that this order should execute in its entirety, or not at all. \\
        Hidden & Boolean, static & Indicates that this order should not be displayed if placed in an order book. \\
        ISO & Boolean, static & Indicates that this order is part of an inter-market sweep, and that standard execution price protections are waived. \\
        Time in Force & Duration, static; $\mu$s & Amount of time that this order should rest in a limit order book before it is cancelled by the exchange. \\
    \end{tabularx}
    \label{tab:add_sv}
\end{table*}

\begin{table*}[htp]
    \centering
    \caption{State variables for the Modify (Mod) message entity.}
    \begin{tabularx}{\textwidth}{llX}
        Variable Name & Variable Type and Units & Meaning \\
        \hline
        Message Header & N/A & See Table~\ref{tab:message_header_sv} for details. \\
        Sequence Number & Integer, static & The Sequence Number of a resting order. \\
        Side & Categorical, static & The Side (bid or offer) of the resting order. \\
        Shares to Remove & Integer, static & Quantity of shares to be removed from the resting order. \\
    \end{tabularx}
    \label{tab:mod_sv}
\end{table*}

\begin{table*}[htp]
    \centering
    \caption{State variables for the Trade message entity.}
    \begin{tabularx}{\textwidth}{llX}
        Variable Name & Variable Type and Units & Meaning \\
        \hline
        Message Header & N/A & See Table~\ref{tab:message_header_sv} for details. \\
        Price & Integer, static; \$0.01 & Execution price of the trade. \\
        Shares & Integer, static & Quantity of shares exchanged. \\
        Triggering Side & Categorical, static & Side of the active order. \\
        ISO & Boolean, static & Whether the active order was an Inter-market Sweep. \\
    \end{tabularx}
    \label{tab:trade_sv}
\end{table*}

\begin{table*}[htp]
    \centering
    \caption{State variables for the Quote message entity.}
    \begin{tabularx}{\textwidth}{llX}
        Variable Name & Variable Type and Units & Meaning \\
        \hline
        Message Header & N/A & See Table~\ref{tab:message_header_sv} for details. \\
        Bid Price & Integer, static; \$0.01 & Highest price among bids in an order book. \\
        Bid Shares & Integer, static & Quantity of shares associated with the highest priced bid. \\
        Offer Price & Integer, static; \$0.01 & Lowest price among offers in an order book. \\
        Offer Shares & Integer, static & Quantity of shares associated with the lowest priced offer. \\
    \end{tabularx}
    \label{tab:quote_sv}
\end{table*}

\begin{table*}[htp]
    \centering
    \caption{State variables for the National Best Bid and Offer (NBBO) message entity.}
    \begin{tabularx}{\textwidth}{llX}
        Variable Name & Variable Type and Units & Meaning \\
        \hline
        Quote & N/A & See Table~\ref{tab:quote_sv} for details. \\
        Bid Exchange & String, static & Identifier of the exchange that holds the National Best Bid. \\
        Offer Exchange & String, static & Identifier of the exchange that holds the National Best Offer. \\
    \end{tabularx}
    \label{tab:nbbo_sv}
\end{table*}

\begin{table*}[htp]
    \centering
    \caption{State variables for the Limit Up-Limit Down (LULD) band message entity.}
    \begin{tabularx}{\textwidth}{llX}
        Variable Name & Variable Type and Units & Meaning \\
        \hline
        Message Header & N/A & See Table~\ref{tab:message_header_sv} for details. \\
        Upper Band & Integer, static & Highest eligible trade price. \\
        Lower Band & Integer, static & Lowest eligible trade price. \\
    \end{tabularx}
    \label{tab:luld_band_sv}
\end{table*}

\begin{table*}[htp]
    \centering
    \caption{State variables for the Receipt message entity.}
    \begin{tabularx}{\textwidth}{llX}
        Variable Name & Variable Type and Units & Meaning \\
        \hline
        Message Header & N/A & See Table~\ref{tab:message_header_sv} for details. \\
        Success & Boolean, static & Indicates whether a request was successful or not. \\
        Reason & String, static & Optional error message indicating why a request failed. \\
    \end{tabularx}
    \label{tab:receipt_sv}
\end{table*}

\begin{table*}[htp]
    \centering
    \caption{State variables for the Trigger message entity.}
    \begin{tabularx}{\textwidth}{llX}
        Variable Name & Variable Type and Units & Meaning \\
        \hline
        Message Header & N/A & See Table~\ref{tab:message_header_sv} for details. \\
        Trigger Event & Categorical, static & What event should be triggered, Auction or Trade. \\
    \end{tabularx}
    \label{tab:trigger_sv}
\end{table*}

\subsection{Scales}\label{subsec:scales}
The minimum observable time increment of ABMMS is 1 $\mu$s, though the simulation ``step size'' is variable and based on scheduled events.
The model is usually run in segments of 1 trading day (6.5 hours), 5 trading days (1 trading week), or 20 trading days (1 trading month).
Spatial relationships are not explicitly represented, though they appear implicitly in the CN, where propagation delays are estimated based on properties of fiber optic communication technology and geographic locations of real world data centers~\cite{tivnan2020fragmentation}.
The round lot size is 100 shares, and is used to filter quotes when constructing an NBBO, but odd lots are not restricted.
The minimum tick size for prices of quotes is \$0.01, trades can occur in \$0.001 increments, and maker-taker fees are in increments of \$0.0001.
% \todo{CVO: Nit pick, the truncated midpoint price implementation makes the trade increment \$0.01.}

All scales present in ABMMS are selected with the intent to model the NMS as closely as possible.
The round lot size and minimum price increments are directly taken from regulation and documentation of NMS participants.
The most subjective choice is the minimum time increment of 1 $\mu$s, which allows for accurate modeling of most trading processes.
However, if agent response times were to be accurately modeled, a smaller minimum increment (i.e.\ 1 ns) may be needed, since exchanges and some high frequency trading strategies may have faster response times than 1 $\mu$s.

\subsection{Process overview}\label{subsec:process_overview}
ABMMS is event-driven, with discrete events occurring in fine-grained, but discrete time.
To initiate a simulation, a start method is called for each agent, allowing it to perform setup actions at run time and schedule initial trading actions.
The results of these start methods form the seed of the event-driven simulation, where messages are processed in increasing order of time of receipt.
Each agent has its own strategies for how it reacts to particular message types, but generally all state variables are updated asynchronously.
The only variable that is updated synchronously is the global clock, which is shared by all agents in a simulation.
The existence of a single global clock removes the possibility of clock synchronization issues, which is a prevalent and difficult problem in distributed high frequency systems.
Events are processed sequentially until there are no remaining events to process or the simulation termination time is reached.

All agents in ABMMS respond to messages, which indicate state changes in other agents or that a scheduled discrete event should occur.

Agent behaviors fall into one of two classes: planned and reactive.
Planned behaviors are triggered by a message sent from the agent to itself, with the intent of performing a certain action at a specific time.
For example, an agent may decide that it would like to submit an order to an exchange in five minutes, and schedule this planned action using a trigger message.

Reactive behaviors are triggered by messages sent to the agent from elsewhere in the system.
For example, suppose an arbitrage trader receives a new quote from an exchange that indicates a crossed market state.
The arbitrage trader then reacts by sending orders to the two exchanges involved in the cross, intending to be the counterparty to the bid at one exchange and the offer at another.

\subsubsection{CN Processes}\label{subsubsec:CN_processes}
Message routing is determined based on the shortest weighted path identified by Dijkstra's algorithm.
The total weight of that path is a deterministic propagation delay, which is then combined with a small exponential random variable to simulate queuing, latency jitter, and other stochastic delays.
All messages are subjected to a minimum propagation delay, which defaults to 5 $\mu$s and mainly impacts messages passed between agents at the same node in the CN\@.

\subsubsection{Exchange Processes}\label{subsubsec:exchange_processes}
Exchanges implement a trading day schedule such that trading occurs between 9:30am and 4pm on week days (Monday through Friday), excluding US federal holidays.
This is enforced by opening and closing processes, outside of which any submitted requests will be rejected.

The opening and closing procedures for exchanges in ABMMS are relatively simple compared to what is seen in the NMS\@.
Orders are not accepted prior to market open, and continuous trading begins immediately at market open without the occurrence of an opening batch auction.
Similarly, the closing process in ABMMS is simply the termination of continuous trading, unlike the NMS where it is common to execute a closing batch auction.
As part of the closing process, all orders that use the default time in force value or have a time in force value less than 17.5 hours (the duration between the close and next open) are cancelled by the exchange.

During open trading hours, Exchanges will validate any incoming requests to ensure that they are well formed.
Valid messages are passed to the Matching Engine, while ill formed messages result in receipts returned to their senders indicating issues.

Beyond these processes, Exchanges track the current NBBO and LULD bands for each trading symbol, and construct a Direct Best Bid and Offer (DBBO) using quotes observed directly from other exchanges.

\subsubsection{Matching Engine Processes}\label{subsubsec:matching_engine_processes}
Add Requests that are validated by an exchange are handed off to a Matching Engine, which determines if that order will immediately lead to trades or if it will fall to rest in a Order Book.
An incoming bid immediately matches against the highest priority offer in the Order Book if the limit price of the bid is greater than or equal to the limit price of the offer.
A similar, but inverted, relationship holds for an incoming offer matching against resting bids.
A single incoming request may result in multiple trades, depending on the quantity of shares desired by the incoming order and the quantity provided by resting orders.
If there are not enough resting shares to complete an incoming request, then that request will fall to rest in the order book and wait till a counterparty is found.

If ABMMS is configured with multiple exchanges and a trade would occur at a price that is worse than what is displayed by the current NBBO, then the Matching Engine will defer execution of that order and route it to the Exchange that holds the appropriate side of the NBBO (assuming that the Matching Engine is not managed by the indicated Exchange).
In addition to forwarding the remainder of the order to the NBBO holder, the routing exchange will also send a receipt to the sender of the request indicating that routing has occurred.
This process is referred to as trade-through protection, and the mechanism that implements trade-through protection is mandatory order routing.

The results of each trade are emitted in a Trade message via the feed of the Exchange that manages a Matching Engine.
In addition, the pair of agents involved in a trade each receive a receipt that indicates their involvement.

Up to this point we have mainly considered the case of Limit Orders, but, Exchanges in ABMMS also support Market Orders and Midpoint Orders, which function slightly differently.
Limit Orders guarantee price, but not execution, using the limit price specified by the trader.
On the other hand, Market Orders guarantee execution, but not price.
This is implemented in ABMMS as a Limit Order whose limit price is set to the appropriate side of the current LULD band (maximum price for Market bids, minimum price for Market offers).
Midpoint Orders are Limit Orders whose limit price is set to the midpoint of the current NBBO, e.g.\ Best Bid Price + Best Offer Price / 2.
The limit price of Midpoint Orders is updated by the Exchange each time a new NBBO is received, and this limit price adjustment does not impact other aspects of the order (i.e.\ submission time).

Beyond these order types, all Add Requests have a set of modifiers that can be applied to shape how they are executed.
The ``Time in Force'' attribute of an order indicates how long it should be considered valid for once the exchange has received it.
By default, orders are considered valid for a single trading day, and will be cancelled as part of the closing process.
Orders with a ``Time in Force'' of 0 $\mu$s are either executed or cancelled immediately.
The ``All or Nothing'' flag indicates that a request should be executed completely or not at all.
In the current implementation, use of the "All or Nothing" flag implicitly causes the order to have a ``Time in Force'' of 0 $\mu$s.
The ``Intermarket Sweep Order'' (ISO) flag disables trade-through protections for the order, ensuring that it is not routed to another Exchange.
Orders can be marked as ``Hidden'', which inhibits the issuance of any Add or Quote messages based on the entry of the order.

\subsubsection{Order Book Processes}\label{subsubsec:order_book_processes}
Order Books store Add Requests that have been accepted by an Exchange, but have not yet been fulfilled.
These Add Requests are stored in a pair of sorted queues, one for bids and one for offers, where the sort ordering is based on price, visibility, submission time, and a random tie-breaker.
For bids, higher prices result in higher priority, while for offers the opposite is true.
Lower visibility, e.g.\ the use of the hidden modifier, always results in a lower priority.
Older orders have priority over more recently submitted orders.

Every Add Request that falls to rest in the Order Book causes an Add Receipt to be issued to the request sender and and Add message to be issued on the feed of the controlling Exchange.

Mod Requests that have been validated by the Exchange are executed by the Order Book, resulting in a Mod Receipt response to the request sender and a Mod message issued on the feed of the controlling exchange.

Any Request that replaces or modifies the Best Bid or Best Offer results in the construction and issuance of a new Quote message.
Note that hidden orders are not considered when constructing a new Quote.

Additionally, the Order Book implements the price updates for any resting Midpoint Orders on behalf of the Exchange any time the NBBO is updated.

\subsubsection{SIP Processes}\label{subsubsec:sip_processes}
Each SIP provides a suite of services for a set of trading symbols.
For each trading symbol in this set, the SIP will construct and disseminate a Trade and Quote (TAQ) feed, a National Best Bid and Offer (NBBO), and Limit Up-Limit Down (LULD) bands.

The TAQ feed is nearly a forwarding service, but the SIP appends an additional timestamp to each message indicating the time that it was received by the SIP.

A NBBO is constructed by aggregating quotes from multiple exchanges that are simultaneously trading the same symbol.
The bid with the highest limit price across Exchanges is selected for the National Best Bid, and the offer with the lowest limit price is selected for the National Best Bid.
In order for a bid or offer to be considered during the construction of an NBBO it must be at least as large as a round lot, where the default round lot size is 100 shares.

LULD bands are constructed using a rolling window of trades, which covers the past 5 minutes by default.
A reference price is constructed using the simple mean of the trade prices.
The band values are then calculated as a percentage deviation from the reference price.
The default LULD band width is $\pm5\%$ for stocks with a share price greater than \$3.00, $\pm20\%$ for stocks with a share price in the range $\$0.75 \leq x \leq \$3.00$, and $\pm75\%$ for stocks with a share price less than $\$0.75$.
Additionally, the LULD band width doubles during the last 25 minutes of trading on regular trading days.

One element of the LULD system that ABMMS does not implement is trading halts.
In a complete implementation, a trading halt would be triggered if trading occurred outside of the LULD bands, and did not return within 15 seconds.
Full details of the LULD system can be found at \url{https://www.luldplan.com/}.

\subsubsection{Trader Processes}\label{subsubsec:trader_processes}
Here we discuss general processes that are shared by all traders, namely a budget constraint and request tracking.
Details related to trading decisions are covered in Section~\ref{subsubsec:adaptation}.

All traders are subject to a budget constraint that restricts their actions when they do not have enough resources.
Specifically, each trader estimates a portfolio value that includes cash and shares of stock.
The value for the shares of stock are estimated based on the prices displayed by the NBBO and the quantity of shares held by the agent.
If the agent owns a positive quantity of shares, then they must sell those shares to close their position, thus the National Best Bid price is used.
Alternatively, if the agent owns a negative quantity of shares (shorting), then they must buy shares in order to close out the position, thus the National Best Offer price is used.
In either case, the estimated value is the number of shares multiplied by the estimated execution price.
This estimate is extremely coarse and simplified, but can be computed quickly, which is necessary since traders operate at relatively high frequencies.

In order to implement the budget constraint, all traders must track their assets.
The only messages that lead to a change in assets are Trade Receipts, which convert between shares and cash.

All traders also perform basic request tracking.
Mod Requests have an immediate impact if accepted, and thus can be disregarded once their impact has been noted.
On the other hand, Add Requests can land in an order book, waiting until a counterparty is found.
Each trader keeps a copy of all outgoing Add Requests.
That copy is updated if the trader receives any associated Mod or Trade Receipts.
Since Exchanges cancel all orders with insufficient time in force values at the end of a trading day, all traders similarly clear any tracked orders with insufficient time in force values at the end of a trading day.

\subsection{Scheduling}\label{subsec:scheduling}
When an agent receives a message, it may create and send any number of messages to arbitrary recipients.
The receipt time of a new message is based on the time it was sent, along with the shortest path (in terms of total propagation delays) between the location of the sender and the location of the intended recipient.
Messages are processed in order of increasing receipt time.

\subsection{Design Concepts}\label{subsec:design_concepts}
\subsubsection{Basic Principles}\label{subsubsec:principles}
ABMMS is designed to emulate the NMS, with the assumption that capturing more details of mechanics in the target system will improve the ability of an ABFM to evaluate the impacts of system perturbations.
Additionally, ABMMS aims to investigate the impact of more realistic agent behavior in a similar lens.
Human traders are flexible, adaptive, and heterogeneous in ways that are not captured by the relatively simple agents that commonly populate ABFMs.
Both of these principles likely have non-trivial impacts on their own, but we hypothesize that their combination will create synergistic effects.

Detailed market mechanics are captured at the system level, with the CN accounting for the details of communication and the fragmented market that is distributed across that CN, as well as at the agent level, with the implementation of realistic exchanges, matching engines, and order books.

Simple agents, particularly ZI~\cite{gode1993allocative} and ZIP~\cite{cliff1997minimal}, provide a foundation upon which we can develop learning agents.
We use meta-reinforcement learning~\cite{wang2016learning, duan2016rl} to develop learning agents with adaptive strategies.
Beyond adapting to changing market conditions, learning agents can also strategically leverage system elements that are largely unused by the simple traders, such as alternative order types and order execution modifiers.

ABMMS aims to generate data that is indistinguishable from data products offered by genuine market participants.
In particular, the primary output of ABMMS is a comprehensive depth of book data feed, which should support a variety of market data analyses.
One class of analyses that are not supported by this depth of book feed are those that require attribution, which is also a limitation of all commercial data products offered in the wild.
The only agents in the NMS that have access to data with attribution are exchanges, regulators, and similar entities.

\subsubsection{Emergence}\label{subsubsec:emergence}
We designed ABMMS to minimize the amount of imposed or prescribed behaviors, with the intent that almost all results collected from ABMMS would be emergent.
The primary output of ABMMS is a full depth-of-book data feed, which is created entirely through mediated agent interactions.
Since a minimal amount agent behaviors are prescribed or imposed, this generated data feed is a completely emergent result.
The current implementation prescribes the ecology of agents that populates each simulation, however, future work may explore evolutionary game theory approaches that allow the simulation population to develop over time.

\subsubsection{Adaptation}\label{subsubsec:adaptation}
The main adaptive behaviors displayed in ABMMS are the strategies implemented by traders.

ZI traders make trading decisions almost completely randomly.
Add Requests are scheduled to avoid request submissions outside of normal trading hours, and are otherwise based on a uniform delay distribution (details in Section~\ref{subsubsec:trader_wait}).
When submitting an Add Request the ZI trader first selects a trading symbol to target, then a side of the market (bid or offer).
The limit price is randomly selected from a uniform distribution that ranges from \$99.75 to \$100.25 initially, then is updated based on the NBBO.
To create some separation between the price distributions of bids and offers, the limit price distribution for bids is shifted downwards by 25\% of the NBBO spread and the limit price distribution for offers is shifted upwards by the same amount.
Add Request volume is determined by a log normal distribution (details in Section~\ref{subsubsec:trader_volume}).
Finally, the exchange that an Add Request routed to is selected at random.
ZI traders only use limit orders, do not use any execution modifiers, and do not submit mod requests.

MI traders are identical to ZI traders, except that they send their Add Requests to the Exchange that holds the appropriate side of the NBBO (details in Section~\ref{subsubsec:trader_routing}).

ZIP traders very similar to ZI traders, but they develop pricing beliefs based on what prices lead to trades.
The limit price used for a particular order, referred to as the shout price by ~\citet{cliff1997zero}, is constructed based on an internal limit price and a multiplicative profit margin then clipped to remain inside of the current LULD bands.
The internal limit price is drawn from a truncated normal distribution (details in Section~\ref{subsubsec:zip_limit_price}).
The profit margins are randomly initialized, and evolves following the Widrow-Hoff ``delta rule'' as discussed in ~\citet{cliff1997zero}.

We verified our ZIP trader implementation by comparison with the implementation provided by ~\citet{cliff2018bse}.
The only notable deviation is the endogenously defined and updated limit prices.

Arbitrage traders construct a synthetic NBBO using direct feeds from each exchange, resulting in a Direct Best Bid and Offer (DBBO).
Each time the Arbitrage trader observes a crossed DBBO, i.e.\ where the best bid price is higher than the best offer price, it emits a pair Add Requests to arbitrage away the cross.
An offer is sent to the Exchange that holds the best bid, and a bid is sent to the Exchange that holds the best offer.
The limit price of the emitted Add Requests is set using the limit price of the order that it is targeting.
Both Add Requests are flagged as ISOs and use a time in force attribute of 0 $\mu$s (i.e.\ immediate or cancel).

The Reinforcement Learning Trader is trained using the IMPortance weighted Actor Learner Architecture (IMPALA) algorithm~\cite{espeholt2018impala}, as implemented by RLLib~\cite{liang2017rllib}.
The IMPALA configuration is summarized in Table~\ref{tab:rl_config}.
The Reinforcement Learning Trader is configured following the Meta-Reinforcement Learning paradigm~\cite{duan2016rl, wang2016learning}, where the policy is memory augmented (using a LSTM~\cite{hochreiter1997long}), the inputs are augmented with the previous action and reward, and the agent is presented with a distribution of tasks during training (many independent instances of ABMMS).

During training the agent learns how to adapt to different market conditions by accounting for the temporal relationships between observations, actions, and rewards.
This adaptation mechanism can be thought of as an inner reinforcement learning algorithm that is implemented by the LSTM.
During evaluation, the agent is no longer updated following the IMPALA training algorithm, but the adaptation strategy learned by the LSTM remains active.

\begin{table*}
    \centering
    \caption{
        Training configuration for the Reinforcement Learning Trader under the IMPALA algorithm.
    }
    \begin{tabular}{lc}
        Label & Value \\
        \hline
        env & ABMMS \\
        rollout\_fragment\_length & 128 \\
        train\_batch\_size & 1024 \\
        horizon & 128 \\
        soft\_horizon & True \\
        lr & 6e-4 \\
        gamma & 0.995 \\
        grad\_clip & 40 \\
        vf\_loss\_coeff & 1e-5 \\
        l2\_coeff & 1e-10 \\
        entropy\_coeff & 1e-10 \\
        normalize\_actions & False \\
        clip\_actions & False \\
        framework & tf2 \\
        model: use\_lstm & True \\
        model: max\_seq\_len & 64 \\
        model: lstm\_use\_prev\_reward & True \\
        model: lstm\_use\_prev\_action & True \\
        model: lstm\_cell\_size & 256 \\
        model: vf\_share\_layers & False \\
        tune.run: run\_or\_experiment & IMPALA \\
        tune.run: time\_budget\_s & 24 hours \\
        tune.run: checkpoint\_freq & 10 \\
        tune.run: checkpoint\_at\_end & True \\
        tune.run: keep\_checkpoints\_num & 10 \\
        tune.run: checkpoint\_score\_attr & custom\_metrics/log\_return\_mean \\
    \end{tabular}
    \label{tab:rl_config}
\end{table*}

\subsubsection{Objectives}\label{subsubsec:objectives}
The Reinforcement Learning Trader optimizes the log returns of its estimated total portfolio value over episodes consisting of 2000 interactions.

ZIP traders minimize the distance between their shout prices and target prices using a Widrow-Hoff ``delta rule'' update, but do not directly optimize for profits.

\subsubsection{Learning}\label{subsubsec:learning}
As mentioned in Sections~\ref{subsubsec:adaptation} and ~\ref{subsubsec:objectives}, the Reinforcement Learning Trader uses meta-reinforcement learning, implemented via the IMPALA algorithm, to optimize its portfolio log returns.
This reinforcement learning policy reacts to incoming messages, and thus learns a direct behavior mapping from observations to actions.
This policy is represented by an artificial neural network constructed with dense layers and an LSTM layer.

The primary goal for the use of learning in the Reinforcement Learning Trader is to develop a free form strategy that is able to adapt to changes in market conditions on the fly, even if those conditions were not observed during training.
This kind of generalization is difficult to achieve with traditional reinforcement learning techniques, which is why we opted to use meta-reinforcement learning.

\subsubsection{Prediction}\label{subsubsec:prediction}
The IMPALA algorithm used to train our RL Trader features implicit prediction through the use of the policy gradient, and explicit prediction through the use of a policy critic that predicts the rewards associated with a sequence of actions.

No other elements of ABMMS explicitly perform prediction as a part of their function.

\subsubsection{Sensing}\label{subsubsec:sensing}
Agents in ABMMS only have direct access to, and full knowledge of, their own state variables.
Messages sent via the CN can communicate the value of internal state variables to other agents, however, due to propagation delays the actual value of a state variable may have already changed by the time other agents receive such a message.

Each time a ZI trader is activated to trade, the following information is available:
\begin{itemize}
    \item Current values for the traders limit price distribution (based of the current NBBO).
    \item Current holdings (used for budget constraint only)
\end{itemize}

MI traders use the same information as ZI traders, but additionally use the current NBBO to determine where to route their Add Request.

Each time a ZIP trader is activated to trade, the following information is available:
\begin{itemize}
    \item Current limit prices
    \item Current profit margins
    \item Current LULD bands (used to clip shout prices)
    \item Current holdings (used for budget constraint only)
\end{itemize}

Each time the RL trader is activated to trade, it responds based on the following:
\begin{itemize}
    % \item Incoming message
    \item Current holdings
    \item DBBO for each trading symbol
    \item The last response emitted
    \item The last reward received
    \item A state vector that can contain information from previous observations of the above elements
    % \item The set of active Add Requests issued by the RL trader
\end{itemize}

\subsubsection{Interaction}\label{subsubsec:interaction}
All interactions in ABMMS are mediated by the CN, which controls the propagation delay for messages sent by any agent.

Traders send messages to, and receive messages from, Exchanges.
SIPs receive messages from Exchanges, then send messages to Exchanges and Traders.
Exchanges may send messages to other exchanges.
Traders do not send messages to other traders.
Note that these are conventions, and are not explicitly enforced.

\subsubsection{Stochasticity}\label{subsubsec:stochasticity}
Stochasticity is used heavily in the initialization of ABMMS, with many of the details discussed below in Section~\ref{subsec:init}.
The primary goal of stochastic initialization in ABMMS is to aid in exploring a neighborhood of similar markets, since any particular initialization is unlikely to match a historical state of the NMS\@.
In the context of evaluating the impacts of policy and design perturbations, it is important to know if the impacts of a particular policy are sensitive to initial conditions.

In the CN, exponentially distributed noise is added to deterministic propagation delays, simulating stochastic elements of electronic communication technology, like queuing delays and packet loss.

Trading decisions made by ZI, MI, and ZIP agents are largely stochastic.
This is important to provide a spark that starts trading in ABMMS\@.
Many classes of hand coded trading strategies are purely reactive, they only respond to take advantage of specific market states.
For example, the Arbitrage Trader waits for a crossed market state and then attempts to profit from uncrossing it.
However, if all traders were purely reactive, then a deadlock would occur at the start of the simulation, with each agent waiting for something to happen, and no trading would occur.
The stochastic actions of ZI, MI, and ZIP agents ensure that this deadlock does not occur, and allows reactive strategies to function in ABMMS\@.

\subsubsection{Collectives}\label{subsubsec:collectives}
ABMMS does not explicitly represent any collectives, though herding behavior may emerge.

\subsubsection{Observation}\label{subsubsec:observation}
The data collected from ABMMS is designed to mimic the features that might be found in a genuine consolidated data product.
That means comprehensive coverage of data feeds from all exchanges and SIPs, full depth of book information, and no attribution data.
A single observer, located at the Carteret node of the CN, is used consistently across all simulation runs to promote comparison.
Selection of the Carteret node is arbitrary, and any of the other CN nodes would serve equally well as the location for the observer, as long as the same location was used across all simulations.
An argument could be made for Secaucus as the location for the observer, since it has the shortest average propagation delay.
However, we chose Cartert to create a more direct comparison with the dislocations observed in ~\citet{tivnan2020fragmentation}.

\subsection{Initialization}\label{subsec:init}
The first step of initialization creates the CN that will mediate all agent communication.
An CN is composed of a network topology, a minimimum propagation delay, and a delay jitter distribution.
The network topology is a weighted and undirected graph that defines the places where agents can be located as well as deterministic propagation delays between those locations.
The minimum propagation delay ensures that all inter-agent communication experiences some amount of delay, avoiding unrealistic scenarios that could occur if some kinds of communication experienced no delays.
The delay jitter distribution adds a stochastic element to the otherwise deterministic communication delays, and is intended to model queuing delays and other stochastic delays present in electronic communication systems.

The default topology closely mimics our understanding of the current communication infrastructure in the NMS, featuing four locations, with Mahwah, Cartert, and Secaucus connected in a triangle, and Weehawken connected Secaucus.
The default minimum propagation delay is 5 $\mu$s, and is based on our understanding of intra-data center communication latency.
The default delay jitter distribution is an exponential distribution with a mean of 5 $\mu$s, and was chosen arbitrarily.

Next, the active trading symbols for the simulation must be defined.
This is deterministic by default, and the specific names used are arbitrary since there are no fundamental value signal or other aspects of real world companies associated with the trading symbols in ABMMS\@.
We use two trading symbols in the default configuration of ABMMS, so that each of the two SIPs present in the default agent configuration manage a single trading symbol.

After the trading symbols are determined, then the various agent groups, Exchanges, SIPs, Traders, and observers, are initialized.
ABMMS currently implements two exchange configurations, one based on the NMS in early 2021 and a simplified system that features a single exchange.
The primary configuration parameters for exchanges, beyond an identifier and location, are a maker fee and a taker fee.
The maker fee and taker fee define a simplified access fee schedule, and allows for the implementation of a traditional maker-taker system as well as an inverted taker-maker system.

In the NMS exchange configuration there are two SIPs, one located in Mahwah and one located in Carteret.
By convention the trading symbols are separated into three groups, Tape A, Tape B, and Tape C.
Trading symbols on Tape A and B are managed by the SIP in Mahwah, while the trading symbols on Tape C are managed by the SIP in Carteret.
In the simplified exchange configuration, there is a single SIP that is colocated with the single exchange and handles all of the trading symbols.

For the configuration of trading agents, the first concern is the ecology of agent types.
ABMMS implements 5 trader types: Zero Intelligence, Minimum Intelligence, Zero Intelligence Plus, Arbitrage, and Reinforcement Learning.
After the ecology of agents has been determined, they must be place at node in the CN\@.
ZI, MI, and ZIP traders are usually placed randomly.
In the currently implemented agent configurations, either no arbitrage traders are included, or a single arbitrage trader is placed at Secaucus.
Similarly, most configurations do not feature any RL agents, but our primary experimental configuration features a single RL agent located at Secaucus.
Each trader is randomly assigned an initial allocation of cash and shares of stock, see Table~\ref{tab:agent_holding_init} for details.

\begin{table*}
    \centering
    \caption{
        Distributions used to initialize trading agent holdings.
        Initial holdings for each trading symbol are drawn independently from the indicated distributions.
        These initial holding distributions are arbitrary.
        Exponential distributions are used based on the understanding that wealth distributions tend to be heavy tailed.
        We chose not to use distributions with unbounded mean and/or variance to improve the consistency of ABMMS results.
    }
    \begin{tabular}{lll}
        Agent Type & Cash Initialization & Stock Initialization \\
        \hline
        RL Trader & Exponential with a mean of \$100 million & 0 shares \\
        Arbitrage Trader & Exponential with a mean of \$100 million & 0 shares \\
        ZIP Trader & Exponential with a mean of \$100 thousand & Exponential with a mean of 10000 shares \\
        MI Trader & Exponential with a mean of \$10 thousand & Exponential with a mean of 10000 shares \\
        ZI Trader & Exponential with a mean of \$10 thousand & Exponential with a mean of 1000 shares \\
        Exchange & \$0 & 0 Shares \\
        Default & Exponential with a mean of \$1 thousand & Exponential with a mean of 100 shares \\
    \end{tabular}
    \label{tab:agent_holding_init}
\end{table*}

The last agents to be initialized are the observers.
The main initialization concern with observers is their number and location.
In all configurations we use a single observer located at the Carteret node.

Once the CN and agents have been initialized, they must be wired together before trading can begin.
All agents subscribe to data feeds provided by agents that they plan to interact with.
The default configuration is that SIPs subscribe to Exchanges, while Exchanges, Traders, and Observers subscribe to Exchanges and SIPs.
Next, all agents are registered with the CN based on their location within the network topology.

To provide a common starting point for trader price beliefs, each SIP issues LULD bands for the trading symbols that it is responsible for.
Since ABMMS does not currently implement opening auctions, the LULD bands issued at the open are based on the closing from the ``previous day'', which is drawn from a uniform distribution ranging from \$90.00 to \$110.00.
Individual agents may initialize their price beliefs based on these LULD bands, or use their own initialization schemes.

ZI and MI traders do not have any parameters beyond their minimum and maximum limit prices that need to be initialized.
Arbitrage traders also do not have any additional parameters that require initialization.

Our ZIP agents follow the same initialization process as ~\citet{cliff1997minimal}.
Specifically, the learning rate for each agent is drawn uniformly from 0.1 to 0.5, the momentum parameter is drawn uniformly from 0.0 to 0.1, and the profit margins are drawn uniformly from 0.05 to 0.35 (values are positive for offer profits and negative for bid profits).
For the target price perturbation functions, absolute shifts are drawn uniformly from \$0.00 to \$0.05 and relative shifts are also uniformly random with a maximum change of plus or minus 5\%.

\subsection{Input Data}\label{subsec:input_data}
ABMMS does not use exogenous input data to represent time-varying processes.

\subsection{Submodels}\label{subsec:submodels}
\subsubsection{Trader Wait Intervals}\label{subsubsec:trader_wait}
ZI, MI, and ZIP traders schedule their trading actions based on a uniform distribution with a minimum wait time of 0.5 seconds and a maximum wait time of 1.5 seconds.

\subsubsection{Trader Order Volume}\label{subsubsec:trader_volume}
ZI, MI, and ZIP traders determine the amount of shares associated with Add Requests based on a log normal distribution with $\mu = 2.6051702,\sigma = 1.40943376$.
These parameters were selected so that the mode of the distribution was close to 100 (the round lot size), and the mean of the distribution was close to 270 (Average Order Size displayed by \url{https://iextrading.com/stats/} on 2021/04/05).

\subsubsection{Trader Order Routing}\label{subsubsec:trader_routing}
ZI and ZIP traders uniformly randomly select the target exchange for each request.

MI traders send their orders to the exchange that holds the appropriate side of the NBBO.
If that side of the NBBO is currently undefined, then MI traders will fall back to random selection.

\subsubsection{ZIP Trader Limit Price}\label{subsubsec:zip_limit_price}
Traditionally, ZIP traders develop shout prices, the price applied to Add Requests, using an exogenous limit price and a multiplicative profit margin.
To avoid providing exogenous limit prices, our ZIP trader implementation instead selects random limit prices following a truncated normal distribution with parameters $\mu = \texttt{Mid}, \sigma = \texttt{Range} / 12.0$, where \texttt{Mid} is the midpoint of the current LULD band and \texttt{Range} is the difference between the upper and lower values of the current LULD band.
The values drawn from this normal distribution are truncated to remain within the current LULD band.
These limit prices are resampled each time a new LULD band is issued.
The denominator used to construct $\sigma$ is arbitrary, and is intended to keep the majority of limit prices centrally located.

\end{document}